\documentclass[prb,twocolumn,showpacs,nofootinbib,preprintnumbers,amsmath,amssymb,aps,longbibliography,superscriptaddress]{revtex4-2}
\usepackage[T1]{fontenc}
\usepackage[utf8]{inputenc}
\usepackage{subfigure, lmodern, amsmath,amssymb, graphicx, pifont, adjustbox, bm, xcolor}
\usepackage{amsfonts}
\usepackage{amsthm}
\usepackage{comment}
\usepackage{mathtools}
\usepackage{float}
\usepackage{braket}
\usepackage{longtable}
\usepackage{graphicx, subfigure}
\usepackage{xcolor}
\usepackage{tikz}
\usepackage{enumitem}
\usepackage{dsfont}
\usepackage{siunitx}
\usepackage{braket}
\usetikzlibrary{decorations.markings}
\usepackage[colorlinks=true, citecolor=blue, urlcolor=blue, linkcolor=blue, breaklinks=true, pdfpagelabels=false]{hyperref}
\newcommand{\tr}{\text{tr}}
\newcommand{\boldr}{\bold{r}}
\newcommand{\boldR}{\bold{R}}
\newcommand{\boldk}{\bold{k}}
\newcommand{\boldtau}{\boldsymbol{\tau}}
\newcommand{\BZ}{|\text{BZ}|}
\newcommand{\boldK}{\bold{K}}
\newcommand{\boldM}{\bold{M}}
\newcommand{\boldsigma}{\boldsymbol{\sigma}}
\newcommand{\boldp}{\bold{p}}
\newcommand{\boldGamma}{\boldsymbol{\Gamma}}
\newcommand{\bolds}{\bold{s}}
\newcommand{\boldq}{\bold{q}}

\newcommand{\boldG}{\bold{G}}

\usepackage{cleveref}
\crefname{appendix}{App.}{Apps.}
\crefname{equation}{Eq.}{Eqs.}
\crefname{figure}{Fig.}{Figs.}
\crefname{table}{Tab.}{Tabs.}
\crefname{section}{Sec.}{Secs.}

\begin{document}
\title{Intervalley-Coupled Twisted Bilayer Graphene from Substrate Commensuration}
\date{\today}
\author{Bo-Ting Chen}
\affiliation{Department of Physics, Princeton University, Princeton, New Jersey 08544, USA}
\author{Michael G. Scheer}
\thanks{Present address: Department of Physics, Harvard University, Cambridge, Massachusetts, 02138, USA}
\affiliation{Department of Physics, Princeton University, Princeton, New Jersey 08544, USA}
\author{Biao Lian}
\affiliation{Department of Physics, Princeton University, Princeton, New Jersey 08544, USA}

\begin{abstract}
We show that intervalley coupling can be induced in twisted bilayer graphene (TBG) by aligning the bottom graphene layer with either of two types of commensurate insulating triangular Bravais lattice substrate. The intervalley coupling folds the $\pm K$ valleys of TBG to the $\Gamma$-point and hybridizes the original TBG flat bands into a four-band model equivalent to the $p_x$-$p_y$ orbital honeycomb lattice model, in which the second conduction and valence bands have quadratic band touchings and can become flat due to geometric frustration. The spin-orbit coupling from the substrate opens gaps between the bands, yielding topological bands with spin Chern numbers $\mathcal{C}$ up to $\pm 4$. For realistic substrate potential strengths, the minimal bandwidths of the hybridized flat bands are still achieved around the TBG magic angle $\theta_M=1.05^\circ$, and their quantum metrics are nearly ideal. We identify two candidate substrate materials Sb$_2$Te$_3$ and GeSb$_2$Te$_4$, which nearly perfectly realize the commensurate lattice constant ratio of $\sqrt{3}$ with graphene. These systems provide a promising platform for exploring strongly correlated topological states driven by geometric frustration.
\end{abstract}

\maketitle

\emph{Introduction.}---Twisted bilayer graphene (TBG) at the magic angle $\theta_M\approx 1.05^\circ$ \cite{Bistritzer2011} has attracted extensive interest in recent years. This system hosts topological flat electron bands with strong interactions \cite{Po2019prb-faithful,Ahn2019prx-failure,Song2019prl-all,TBG2} and possesses a rich phase diagram with superconductivity, correlated insulator, and Chern insulator phases \cite{Cao_2018_Correlated,Cao_2018_Unconventional,Yankowitz_2019,Chen_2019,Lu_2019,Choi_2019,Sharpe_2019,Polshyn2019,Codecido2019,Liu_2020,Shen_2020,Chen2020,Andrei_2020,Serlin_2020,Nuckolls_2020,Xie_2021,TBG4,Bultinck_2020,Kwan_2021,Nuckolls_2023}. The exploration of flat bands in various other 2D moir\'e systems has also led to fruitful discoveries of novel correlated and topological states such as the fractional Chern insulators (FCIs) at zero magnetic field recently observed in twisted MoTe$_2$ \cite{Redekop2024,Cai_2023,Zeng:2023rxr,Park_2023,Xu_2023,Ji_2024,Kang2024,xu2024,park2024ferromagnetism} and pentalayer rhombohedral graphene \cite{Lu2024}. Generically, the topological and geometrical properties of flat bands are crucial for realizing strongly correlated and topological states such as FCIs \cite{Tarnopolsky_2019,WangJie2021prl,Claassen2015prl_position,Peotta_2015,Xie_2020_prl_topology}, and moir\'e systems are an ideal platform for tuning these properties.

In this letter, we are interested in designing graphene-based moir\'e flat bands with a geometric frustration origin, such as the flat bands of the kagome lattice and $p_x$-$p_y$ orbital honeycomb lattice tight-binding models \cite{Lieb1989prl_two,A_Mielke_1991,Dumitru2022naturephysics,CongjunWu2007,Bergman_2008}, which are promising systems for spin liquids and FCIs \cite{Yan_2011,Yin_2022,Tang_2011}. While such moir\'e flat bands were previously predicted in $\Gamma$-valley twisted transition metal dichalcogenides (TMDs) \cite{Angeli_2021,Xian_2021,Liu_2022,wang2022moireengineering}, these predicted bands lie far from charge neutrality and may suffer from inaccuracies of TMD model parameters. Recently, it was shown that twisted Kekul\'e ordered graphene systems can give kagome lattice and $p_x$-$p_y$ orbital honeycomb lattice flat bands \cite{scheer2023prl,scheer2023,scheer2022} due to the intervalley coupling induced by Kekul\'e order. Kekul\'e ordered graphene can be synthesized by lithium deposition \cite{Sugawara_2011,Kohei_2012_Ca,Bao_2021,Cheianov_2009_Hidden,Qu:2022sev}, but this method introduces disorder and electron doping and is challenging in the twistronics context.

This motivates us to study the engineering of the TBG flat bands through intervalley coupling arising from a substrate material. We show that an intervalley coupling can be induced in TBG by aligning the bottom graphene layer with an insulating substrate of either of two types of commensurate lattice. This eliminates the valley degree of freedom and modifies the magic angle TBG flat bands into a $p_x$-$p_y$ orbital honeycomb lattice model which hosts flat bands with topological quadratic band touchings \cite{CongjunWu2007}. With spin-orbit coupling (SOC) from the substrate, these flat bands develop into flat spin Chern bands with spin Chern number up to $\pm4$ and close-to-ideal quantum metrics \cite{WangJie2021prl,Claassen2015prl_position}. From \emph{ab initio} calculations, we identify two candidate substrate materials Sb$_2$Te$_3$ and GeSb$_2$Te$_4$. Intervalley-coupled TBG provides a new platform for studying interacting topological states in geometrically frustrated flat bands.

\emph{The model setup.}---We consider a TBG system on a commensurate \emph{non-magnetic} substrate as follows. The top and bottom graphene layers (denoted by $l=\pm$) are twisted relative to an aligned configuration by angles $-l\theta/2$ so that the two layers have a relative twist angle of $\theta$. The bottom graphene layer is aligned with a substrate with a triangular Bravais lattice commensurate with the graphene lattice. We require the substrate to be a gapped insulator with the graphene chemical potential lying in the gap. This way, the substrate contributes no electrons to graphene and simply induces a substrate potential at low energies. This gives a generic Hamiltonian for the TBG system of the form
\begin{equation}
H = 
\begin{pmatrix}
\mathcal{H}_{+} & \mathcal{H}_{\text{hop}} \\
\mathcal{H}_{\text{hop}}^{\dagger} & \mathcal{H}_{-}
\end{pmatrix}
,\quad
\begin{cases}
&\mathcal{H}_{+}=\mathcal{H}_{+}^{(0)}\ ,\\
&\mathcal{H}_{-}=\mathcal{H}_{-}^{(0)}+\mathcal{H}_{-}^{\text{(sub)}}\ ,
\end{cases}
\end{equation}
where $\mathcal{H}_{l}^{(0)}$ is the graphene Hamiltonian in layer $l$, $\mathcal{H}_{-}^{\text{(sub)}}$ is the substrate potential acting on layer $l=-$, and $\mathcal{H}_{\text{hop}}$ is the hopping between the two graphene layers.

Each graphene layer $l=\pm$ has Dirac electrons at two valleys $\boldK_l$ and $-\boldK_l$ in the graphene Brillouin zone (BZ). In this work, we are interested in substrates which couple the two graphene valleys of layer $l = -$. This constrains the substrate lattice constant, as we will show below.

\begin{figure}[tbp]
\centering
\includegraphics[width=80mm]{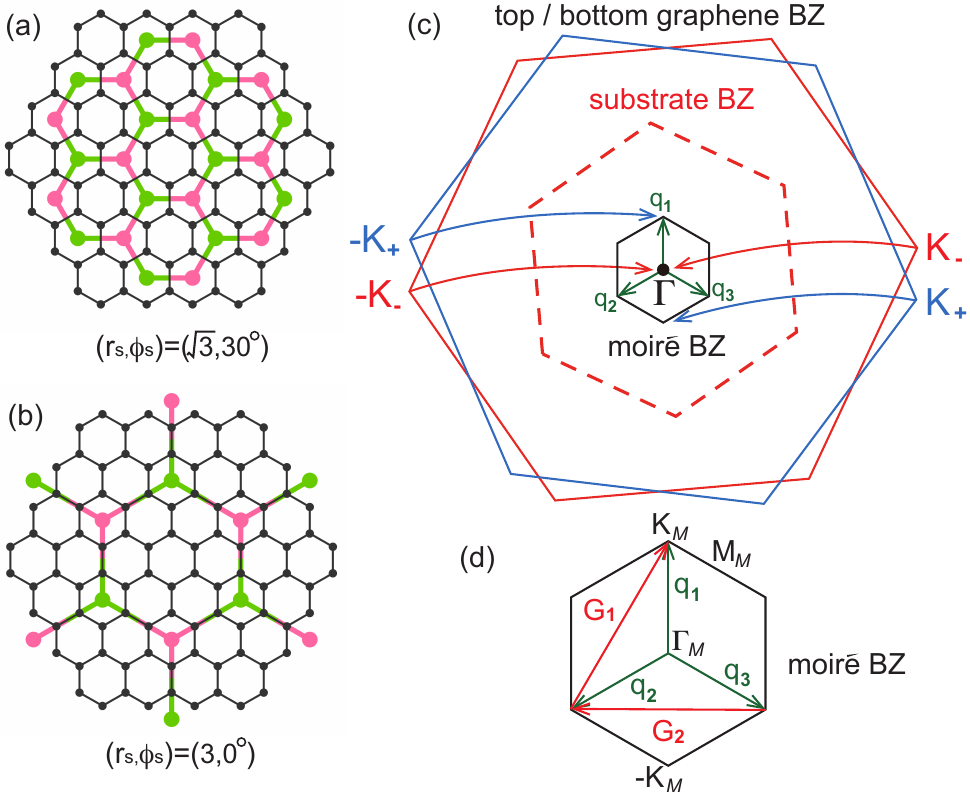}
\caption{(a)-(b): Top view of the bottom graphene lattice (black) on a substrate lattice (pink and green), with $(r_s,\phi_s)=(\sqrt{3}, 30^\circ)$ (type Y) in (a), and $(r_s,\phi_s)=(3, 0^\circ)$ (type X) in (b). (c) The BZs of the top/bottom layer graphene (blue/red solid hexagon) and the substrate in (a) (red dashed hexagon), and the moir\'e BZ (black hexagon). (d) Zoom-in of the moir\'e BZ.}
\label{fig: graphene top view}
\end{figure}

In the continuum limit, we adopt a real space basis $\ket{\boldr,l,\eta,\alpha,s}$, in terms of position $\boldr$, layer $l$, graphene valley $\eta=\pm$ (for $\boldK_l$ and $-\boldK_l$), sublattice $\alpha=\pm$ (for $A$ and $B$), and $z$-direction spin $s=\pm$ (for $\uparrow$ and $\downarrow$). We use $\tau_\mu$, $\sigma_\mu$, and $s_\mu$ to denote the $2\times2$ identity ($\mu=0$) and Pauli ($\mu=x,y,z$) matrices in the basis of valley $\eta$, sublattice $\alpha$, and spin $s$, respectively. The graphene Hamiltonians $\mathcal{H}_{l}^{(0)}$ and $\mathcal{H}_{\text{hop}}$ are spin independent because of the negligible spin-orbit coupling (SOC) in graphene, and they do not couple the two graphene valleys. Explicitly, these terms give the Bistritzer-MacDonald (BM) continuum model \cite{Bistritzer2011} of TBG without substrate:
\begin{equation}\label{eq:H0-Hhop}
\begin{split}
\mathcal{H}_{l}^{(0)} &=-i\hbar v_l\nabla\cdot\Big[\tau_{+}\boldsigma_{l\theta/2}-\tau_{-}\boldsigma_{l\theta/2}^*\Big]s_0, \\
\mathcal{H}_{\text{hop}} &=\Big[\tau_{+}T(\boldr)+\tau_{-}T^*(\boldr)\Big]s_0\ ,
\end{split}
\end{equation}
where $v_l$ is the layer $l$ Fermi velocity, $\tau_{\pm}=\frac{1}{2}(\tau_0\pm\tau_z)$, and $\boldsigma_{\phi} = (\cos(\phi) \sigma_x + \sin(\phi) \sigma_y, -\sin(\phi) \sigma_x + \cos(\phi) \sigma_y)$. We take $v_+=v_-=\SI{611.4}{\milli\electronvolt\nano\meter}$ which is typical \cite{tbg6}, assuming that any substrate effects on $v_l$ are negligible. The periodic moir\'e hopping $T(\boldr)$ is
\begin{equation}
\begin{split}
T(\boldr) &= \sum_{j=1}^{3} T_{\boldq_j} e^{i \boldq_j\cdot \boldr}\ ,\\
T_{\boldq_j} &= w_0 \sigma_0 + w_1 (\sigma_x \cos\zeta_j + \sigma_y \sin\zeta_j)\ ,
\end{split}
\end{equation}
where $\boldq_j = R_{\zeta_j} (\boldK_{-} - \boldK_{+})$ as illustrated in Fig.~\ref{fig: graphene top view}(c)-(d), $R_{\zeta_j}$ represents rotation by angle $\zeta_j=\frac{2\pi}{3}(j-1)$, and the coefficients $w_0$ and $w_1$ represent the interlayer hopping at AA and AB stacking centers, respectively. We set $w_0=88$meV and $w_1=110$meV, which are typical for TBG with $\theta\sim1^\circ$ due to lattice relaxations \cite{Carr_2019}.

The possible commensurate configurations between a triangular Bravais lattice substrate and the bottom graphene layer $l=-$ are labeled by a pair of parameters $(r_s,\phi_s)$, where $r_s=a_s/a_0$ is the ratio between the substrate lattice constant $a_s$ and graphene lattice constant $a_0=\SI{2.46}{\angstrom}$, and $\phi_s$ is the angle between the primitive lattice vectors of the substrate and the bottom graphene layer. We assume $r_s > 1$, since most substrates have lattice constants larger than $a_0$. We also assume the stacking maximizes rotational symmetry, as illustrated in \cref{fig: graphene top view}(a)-(b).

To couple the two valleys of the bottom graphene layer, the commensurate configuration $(r_s,\phi_s)$ is required to fold both $\boldK_-$ and $-\boldK_-$ to the $\boldGamma$ point \cite{scheer2023}, as shown in \cref{fig: graphene top view}(c). A list of such configurations is given in the supplemental material (SM) \cite{suppl}. We focus on two simple types of intervalley coupling configurations:
\begin{equation}\label{eq:comm-para}
\begin{cases}
(r_s,\phi_s)=\left( \frac{\sqrt{3}\rho}{2\mu} , 30^\circ \right)\ , \qquad (\text{Type Y})\\
(r_s,\phi_s)=\left( \frac{3\rho}{\mu} , 0^\circ \right)\ ,  \qquad\quad (\text{Type X})
\end{cases}
\end{equation}
where $\mu$ and $\rho$ are coprime positive integers ($\text{gcd}(\mu, \rho)=1$) and $\mu$ is not divisible by $3$. \cref{fig: graphene top view}(a) shows a type Y configuration with $(r_s,\phi_s)=(\sqrt{3},30^\circ)$. The distortion induced on the graphene lattice by this substrate configuration is known as Kekul\'e-O order \cite{scheer2023prl,Bao_2021}. \cref{fig: graphene top view}(b) shows a type X configuration with $(r_s,\phi_s)=(3,0^\circ)$. In both cases, the black (pink and green) lattice represents the bottom graphene layer (substrate layer).

The folding of the bottom layer $\pm \boldK_-$ points to the $\boldGamma$ point effectively yields a ``$\Gamma$-valley'' TBG moir\'e model, where the top layer $\pm \boldK_+$ points correspond to the $\mp \boldK_M$ points of the moir\'e BZ (\cref{fig: graphene top view}(c)-(d)). Since there is no intralayer coupling between the top layer electrons at $\pm \boldK_+$, the moir\'e BZ has the same size as that of the original TBG system without a substrate \cite{scheer2023prl}. In particular, the moir\'e reciprocal lattice is generated by $\boldq_1 - \boldq_2$ and $\boldq_1 - \boldq_3$, and the Hamiltonian commutes with the translation operators given by
\begin{equation}\label{eq:Trans}
T_\boldR\ket{\boldr, l, \eta, \alpha, s} = e^{i(\boldq_1 \cdot \boldR)\eta(l + 1)/2} \ket{\boldr + \boldR, l, \eta, \alpha, s}
\end{equation}
for $\boldR$ in the moir\'e superlattice. We note that the $T_\boldR$ operators here are different from the translation operators typically chosen for the BM model, which are given by $T^{\text{BM}}_\boldR\ket{\boldr, l, \eta, \alpha, s} = e^{-i l \eta (\boldq_1 \cdot \boldR)}\ket{\boldr + \boldR, l, \eta, \alpha, s}$ (see Table S1 in Ref. \cite{scheer2023prl}). The moir\'e model falls into three symmetry classes as follows.

\emph{$C_{2z}$ symmetric substrates.}---For both types of commensurate configurations in \cref{eq:comm-para}, the maximal \emph{spinful} symmetry the substrate can have consists of the 3-fold rotational symmetry $C_{3z}$, 2-fold rotational symmetry $C_{2z}$, mirror symmetry $M_{\hat{y}}$ which reflects $(x,y,z)$ to $(x,-y,z)$, and time-reversal symmetry $\mathcal{T}$. This maximal symmetry can be achieved for example by a hexagonal lattice with equivalent atoms in the two sublattices (pink and green in \cref{fig: graphene top view}(a)-(b)). Moreover, we make the typical assumption that the substrate induced SOC is momentum independent and spin $s_z$ conserving (for a discussion, see the SM \cite{suppl}). These constraints ensure that the substrate potential takes the generic form:
\begin{equation}\label{eq:max-sub}
\mathcal{H}_{-}^{\text{(sub)}} 
= m_{xxI} \tau_{x} \sigma_{x} s_{0} 
+ m_{zzz} \tau_{z} \sigma_{z} s_{z} 
+ m_{yyz} \tau_{y} \sigma_{y} s_{z} 
,\end{equation}
where $m_{xxI}$ is the spin independent intervalley coupling studied in Ref. \cite{scheer2023prl}, while $m_{zzz}$ and $m_{yyz}$ originate from the substrate intrinsic SOC (see the SM \cite{suppl} for details). For typical substrates, the couplings $m_{xxI}$, $m_{zzz}$ and $m_{yyz}$ are on the order of $\SI{10}{\milli\electronvolt}$. When the two valleys are coupled there is no valley degeneracy. However, $C_{2z}\mathcal{T}$ symmetry forces all the moir\'e bands to be $2$-fold spin degenerate at all momenta. We use integer $n>0$ ($n<0$) to label the $|n|$-th spin-degenerate conduction (valence) moir\'e band relative to charge neutrality.

\begin{figure}[t]
\centering
\includegraphics[width=85mm]{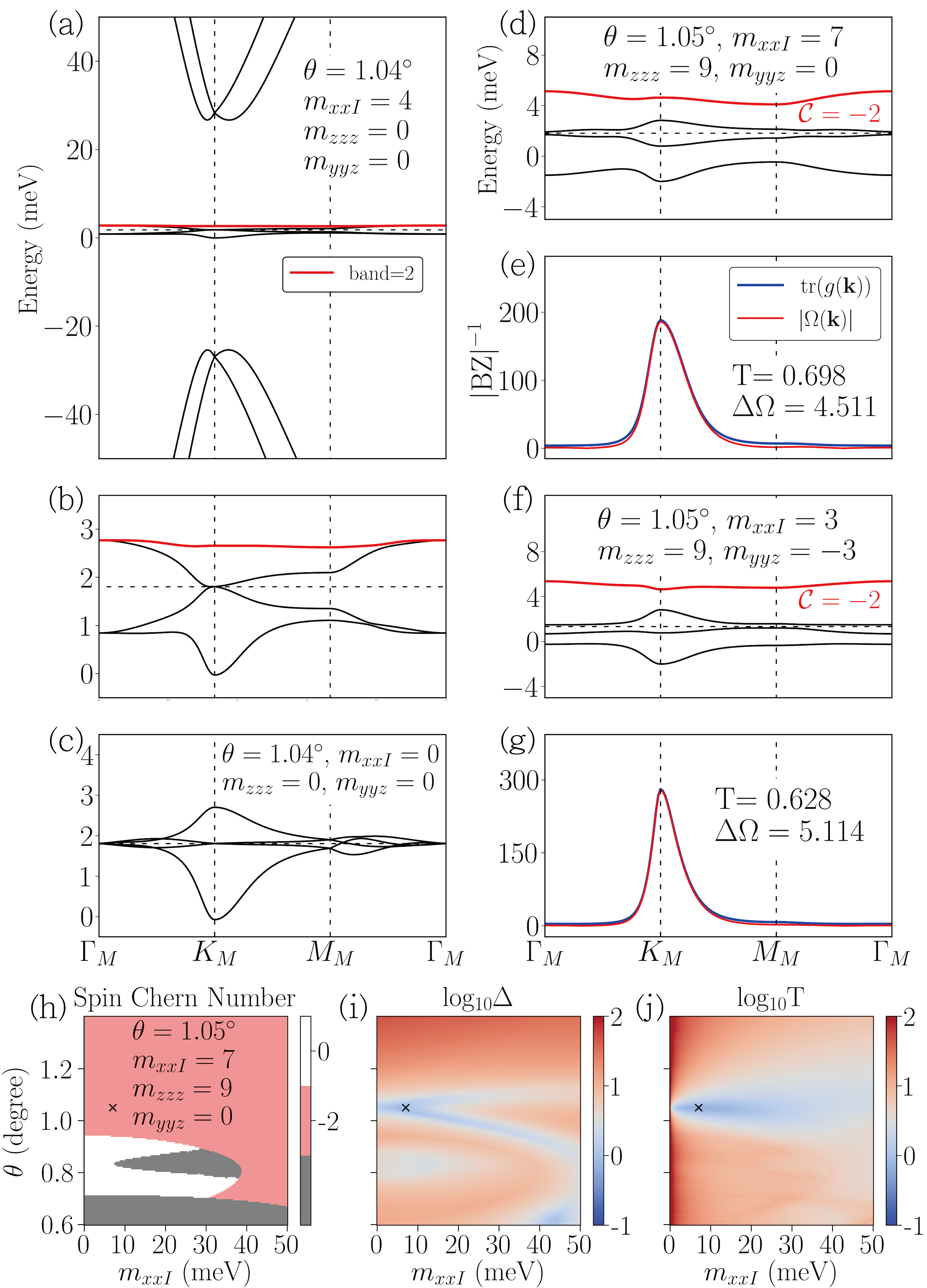}
\caption{The moir\'e band structure for the $C_{2z}$ symmetric substrate potential in \cref{eq:max-sub}. The parameters are given in the panels, and the horizontal dashed line indicates the charge neutrality point. (a) The non-SOC band structure with only $m_{xxI}\neq 0$, for which (b) shows the zoom-in of the $|n|\le 2$ bands. (c) The flat bands (both valleys) of the TBG model without substrate. (d) and (f) are two examples with SOC, which have spin Chern numbers $\mathcal{C}=\{-2, +1, -1, +2\}$ and $\mathcal{C}=\{-2, 0, 0, +2\}$ for bands $n=\{2,1,-1,-2\}$, respectively. (e) and (g) show $\tr(g(\boldk))$ and $|\Omega(\boldk)|$ of the $n=2$ band in (d) and (f), respectively. For $m_{zzz}=9$ and $m_{yyz}=0$, we plot (h) the Chern number (grey denotes regime with almost gapless bands), (i) $\log_{10}\Delta$ where $\Delta$ is the bandwidth, and (j) the $\text{T}$ value of the $n=2$ band with respect to $\theta$ and $m_{xxI}$, where the highlighted point corresponds to parameters in (d)-(e).}
\label{fig: information of all ideal flat bands part1}
\end{figure}

Without SOC, namely $m_{zzz}=m_{yyz}=0$, an example of the moir\'e band structure is shown in \cref{fig: information of all ideal flat bands part1}(a), where charge neutrality is indicated by the horizontal dashed line, and we set $m_{xxI}=4$meV. \cref{fig: information of all ideal flat bands part1}(b) shows the zoom-in of the lowest four ($|n|\le 2$, not counting spin degeneracy) bands around charge neutrality, which originate from intervalley hybridization of the original TBG flat bands (two per valley). The $n=\pm2$ bands have topological quadratic band touching with the $n=\pm1$ bands at $\boldGamma_M$, respectively, carrying $2\pi$ Berry phase. Between the $n=\pm1$ bands, there are two Dirac points at $\pm \boldK_M$. As shown in Ref. \cite{scheer2023prl} which studied the non-SOC model here at large $m_{xxI}$, these lowest four bands are topologically equivalent to the $p_x$-$p_y$ two-orbital tight-binding model in a honeycomb lattice \cite{CongjunWu2007}:
\begin{equation}\label{eq:two-orbital-honeycomb}
H_{\text{tb}}=\sum_{\ell,\ell'=\pm 1} t_{\ell \cdot \ell'} \sum_{\braket{j,j'}} e^{i(\ell-\ell')\varphi_{j',j}} \ket{j',\ell'}\bra{j,\ell}\ ,
\end{equation}
where $\ket{j,\ell}$ is an angular momentum $\ell=\pm1$ orbital on site $j$, $\braket{j,j'}$ denotes nearest neighbors, $t_\pm$ are real hopping parameters, and $\varphi_{j',j}$ is the angle of the vector from site $j$ to site $j'$ relative to some fixed axis. The $n=\pm2$ bands become exactly flat when $|t_+|=|t_-|$ due to geometrical frustration \cite{Dumitru2022naturephysics}. This tight-binding limit can be approached by increasing $m_{xxI}$ and tuning $\theta$ \cite{scheer2023prl}. Here we find the small substrate-induced $m_{xxI}$ can readily make one of the $n=\pm2$ bands extremely flat. For instance, for $m_{xxI}=4$meV in \cref{fig: information of all ideal flat bands part1}(b), the $n=2$ band (highlighted in red) is extremely flat at $\theta=1.04^\circ$.

It is instructive to note that the $\Gamma_M$ point of the moir\'e BZ here in \cref{fig: graphene top view}(d) are the $\pm K_M$ points of the conventionally defined TBG moir\'e BZ of two valleys. Thus, the original magic angle TBG flat bands of two valleys without substrate have a $4$-fold degeneracy (from two Dirac points) at $\Gamma_M$ point of the moir\'e BZ here, as shown in \cref{fig: information of all ideal flat bands part1}(c). This $4$-fold degeneracy is lifted by the substrate intervalley coupling, yielding the four-band model in \cref{eq:two-orbital-honeycomb} and \cref{fig: information of all ideal flat bands part1}(b).

With SOC (nonzero $m_{zzz}$ or $m_{yyz}$), gaps generically open between bands, and each 2-fold degenerate band can carry a spin Chern number $\mathcal{C}=\mathcal{C}_\uparrow =-\mathcal{C}_\downarrow$ due to time-reversal symmetry $\mathcal{T}$ and $s_z$ conservation, where $\mathcal{C}_\uparrow$ ($\mathcal{C}_\downarrow$) is the Chern number of the spin $\uparrow$ ($\downarrow$) band. \cref{fig: information of all ideal flat bands part1}(d) and (f) show two examples of the lowest four bands with SOC terms $m_{zzz}$ and $m_{yyz}$ nonzero (within $\pm \SI{10}{\milli\electronvolt}$) at $\theta=1.05^\circ$, in which at least one of the $n=\pm2$ bands becomes very flat (see also \cref{fig: information of all ideal flat bands part1}(i)). The spin Chern numbers of the $n=\pm2$ bands are robustly $C=\pm 2$ for $\theta>0.95^\circ$, as shown in \cref{fig: information of all ideal flat bands part1}(h). The spin Chern numbers of the $n=\pm1$ bands vary from $0$ up to $\pm4$ as the parameters are varied.

We further examine the quantum geometric tensor (QGT) of the above flat bands, which has recently been found to play an important role in flat band many-body physics, e.g., lower-bounding the superconductor superfluid weight, electron-phonon and optical couplings (see \cite{yu2024quantumgeometryquantummaterials,gao2025} for a review). The QGT for a Bloch band wavefunction $\ket{\psi(\boldk)}$ is defined as \cite{Anandan1990prl, Page1987pra,resta2011} $G_{ij}(\boldk) = 
\tr\left[ 
P(\boldk) \partial_{k_i} P(\boldk) \partial_{k_j} P(\boldk) 
\right]$, where $i,j\in\{x,y\}$, and $P(\boldk)= \ket{\psi(\boldk)} \bra{\psi(\boldk)}$. It can be decomposed into $G_{ij}(\boldk) = g_{ij}(\boldk) + \frac{i}{2} f_{ij}(\boldk)$, where $g_{ij}(\boldk)$ is a real symmetric positive semi-definite matrix known as the \emph{quantum metric}, and $f_{ij}(\boldk)$ is a real antisymmetric matrix giving the Berry curvature $\Omega(\boldk)=-f_{xy}(\boldk)$. They obey an inequality $\tr (g(\boldk)) \ge 2 \sqrt{\det (g(\boldk))}\ge |\Omega(\boldk)|$, and a band with $\tr (g(\boldk))=|\Omega(\boldk)|$ saturating the inequality is called an \emph{ideal} band \cite{WangJie2021prl,Ledwith2023prb,Estienne2023prr,Rahul2014prb}. Particularly, \emph{ideal Chern bands} with Chern number $\mathcal{C}=1$ resemble the lowest Landau level (which has $\tr (g(\boldk))=|\Omega(\boldk)|= \text{Const}$) and allow analytical construction of FCI wavefunctions \cite{Ledwith2022prl,Ledwith2023prb,WangJie2021prl}, which are thus conjectured to be promising platforms for FCIs. In our model, the $n=2$ flat band in \cref{fig: information of all ideal flat bands part1}(d) and (f) are both almost an ideal spin Chern band of Chern number $2$, which can be seen from their $\tr (g(\boldk))$ and $|\Omega(\boldk)|$ (of a particular spin) plotted in \cref{fig: information of all ideal flat bands part1}(e) and (g), respectively. We further define for a band
\begin{equation}
\begin{split}
\label{eqn: definition of T and Delta Omega}
\text{T} &= \frac{1}{2\pi} 
\int_{\text{BZ}} d^2 \boldk 
\left[ \tr(g(\boldk)) - |\Omega(\boldk)| \right]\ ,
\\
(\Delta \Omega)^2 &= \frac{|\text{BZ}|}{4\pi^2}
\int_{\text{BZ}} d^2 \boldk 
\left(
\Omega(\boldk) - \frac{2\pi \mathcal{C}}{|\text{BZ}|}
\right)^2\ .
\end{split}
\end{equation}
which characterize the ideality and Berry curvature uniformity, respectively. 
The values of $\text{T}$ and $\Delta\Omega$ are indicated in \cref{fig: information of all ideal flat bands part1}(e) and (g). The colormap of $\log_{10}(\text{T})$ in \cref{fig: information of all ideal flat bands part1}(j) shows that the $n=2$ band is closest to ideal around $\theta=1.05^\circ$. More colormaps for different parameters are given in the SM \cite{suppl}.

\begin{figure}[t]
\centering
\includegraphics[width=88mm]{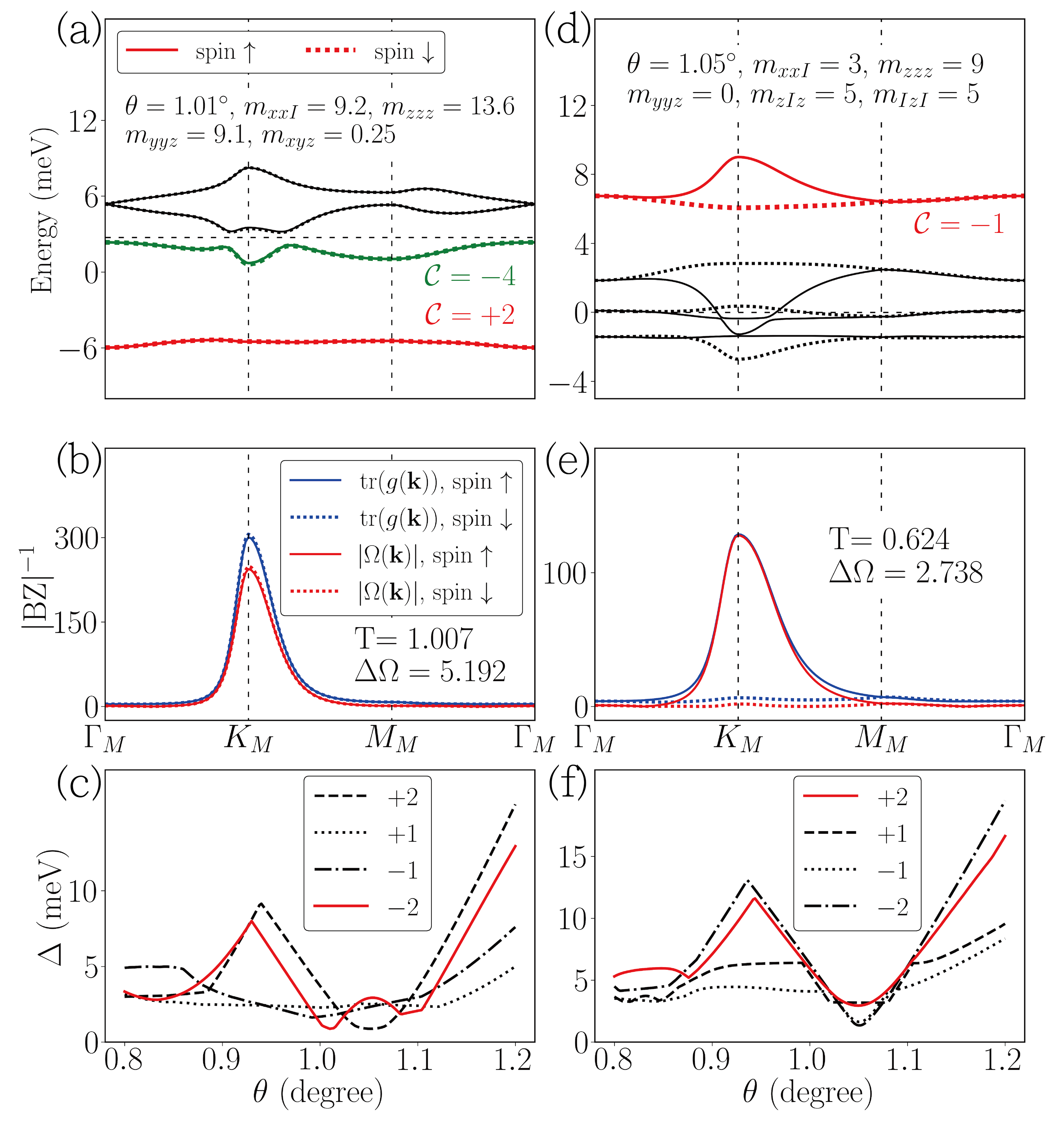}
\caption{(a) and (d) show examples of the moir\'e bands for type Y substrate and type X substrates, respectively, in which the bands $n=\{2,1,-1,-2\}$ have spin Chern numbers $\mathcal{C}=\{-2, +4, -4, +2\}$ and $\mathcal{C}=\{-1, -1, 0, +2\}$, respectively. The parameters in (a) are for Sb$_2$Te$_3$ in \cref{table: candidate substrate sqrt3 concise}. (b) and (e) show $\tr(g(\boldk))$ and $|\Omega(\boldk)|$ for band $n=-2$ in (a) and band $n=2$ in (d), respectively. (c) and (f) show the bandwidths of band $n$ (labeled in the legend) with respect to twist angle $\theta$ for substrate potentials in (a) and (d), respectively.}
\label{fig: information of all ideal flat bands part2}
\end{figure}

\emph{$C_{2z}$ breaking substrates}.---For substrates without $C_{2z}$ symmetry, such as systems with two distinct atomic species on a hexagonal lattice, the bands are no longer forced to have 2-fold spin degeneracy except for the Kramers degeneracy at the time-reversal invariant momenta $\boldGamma_M$ and $R_{\zeta_j}\boldM_M$ for $j \in \{1,2,3\}$. The spin Chern number $\mathcal{C}$ for each pair of bands related by time-reversal is still protected by $s_z$ conservation. We now consider the two types of substrates in \cref{eq:comm-para} separately.

We begin with type Y substrates, which have symmetries $C_{3z}$, $M_{\hat{y}}$, and $\mathcal{T}$, as can be seen in \cref{fig: graphene top view}(a). This constrains the substrate potential (up to unitary transformations preserving $\mathcal{H}_{l}^{(0)}$ and $\mathcal{H}_{\text{hop}}$) to the following form (see SM Section II).
\begin{equation}
\begin{split}
\mathcal{H}_{-}^{\text{(sub)}} 
&= m_{xxI} \tau_{x} \sigma_{x} s_{0} 
+ m_{zzz} \tau_{z} \sigma_{z} s_{z} 
+ m_{yyz} \tau_{y} \sigma_{y} s_{z} \\
&+ m_{xyz} \tau_{x} \sigma_{y} s_{z}\ ,
\end{split}
\end{equation}
where $m_{zzz}$, $m_{yyz}$, and $m_{xyz}$ arise from SOC.

Employing \emph{Quantum Espresso} \cite{Giannozzi_2009, Giannozzi_2017} for first principles calculations, we identify two candidate type Y substrate materials with a $r_s\approx \sqrt{3}$ lattice constant ratio: Sb$_2$Te$_3$ with lattice constant $a_s=4.26 \AA$ ($\frac{r_s}{\sqrt{3}}=0.9998$) \cite{kou2015, Bian_2016, Cao_2016, Jin2013_Proximity}, and GeSb$_2$Te$_4$ with lattice constant $a_s=4.299 \AA$ ($\frac{r_s}{\sqrt{3}}=1.009$) \cite{Talirz_2020,Mounet_2018,Campi_2023}. Both materials are band insulators with the graphene Fermi energy in the band gap. We take the approximation that when graphene is stacked on each of these substrates, the substrate relaxes to exactly realize a commensurate configuration with $r_s = \sqrt{3}$. With this assumption, we determine the substrate potential parameters of these materials (see SM \cite{suppl}) and list them in \cref{table: candidate substrate sqrt3 concise}.

For TBG with monolayer Sb$_2$Te$_3$ substrate (parameters in \cref{table: candidate substrate sqrt3 concise}), \cref{fig: information of all ideal flat bands part2}(a) show the $|n|\le2$ moir\'e bands at $\theta=1.01^\circ$, where solid (dotted) lines stand for spin $\uparrow$ ($\downarrow$) bands. The $|n|\le2$ bands from high to low energies carry spin Chern numbers $\mathcal{C}=\{-2, +4, -4, +2\}$, respectively. Moreover, the $n=-2$ band is extremely flat at $\theta=1.01^\circ$, and its quantum metric is reasonably close to ideal (\cref{fig: information of all ideal flat bands part2}(b)). We also note that the $n=-1$ band is a robust $\mathcal{C}=-4$ flat band within $\theta\in[0.87^\circ,1.05^\circ]$ (see \cite{suppl} Fig. S5 and S7). The results for TBG with GeSb$_2$Te$_4$ substrate is given in the SM \cite{suppl}.

\begin{table}[t]
\centering
\begin{tabular}{|c|c|c|c|c|c|}
\hline
\textbf{Substrate} & {$m_{xxI}$(meV)} &{$m_{zzz}$(meV)} & {$m_{yyz}$(meV)} &{$m_{xyz}$(meV)} \\ \hline
$\text{Sb}_2 \text{Te}_3$ & 9.2 & 13.6 & 9.1 & 0.25 \\ \hline
$\text{Ge} \text{Sb}_2 \text{Te}_4$  & 8.9 & 5.7 & 6.3 & 4.4 \\ \hline
\end{tabular}
\caption{Type Y substrate candidates $\text{Sb}_2 \text{Te}_3$ (lattice constant $4.252 \AA$) and $\text{Ge} \text{Sb}_2 \text{Te}_4$ (lattice constant $4.255 \AA$), which have $r_s\approx \sqrt{3}$. The parameters are fitted from first principles band structure of monolayer graphene on monolayer substrate, with their lattice ratio relaxed to exactly $r_s=\sqrt{3}$.}
\label{table: candidate substrate sqrt3 concise}
\end{table}

We next consider type X substrates, which have symmetries $C_{3z}$, $M_{\hat{x}}$, and $\mathcal{T}$, as can be seen in \cref{fig: graphene top view}(b). In this case, the symmetry constrained substrate potential has the following form (see SM Section II).
\begin{equation}
\begin{split}
\mathcal{H}_{-}^{\text{(sub)}} 
&= m_{xxI} \tau_{x} \sigma_{x} s_{0} 
+ m_{zzz} \tau_{z} \sigma_{z} s_{z} 
+ m_{yyz} \tau_{y} \sigma_{y} s_{z} \\
&+ m_{zIz} \tau_{z} \sigma_{0} s_{z} + m_{IzI} \tau_{0} \sigma_{z} s_{0}\ .
\end{split}
\end{equation}
Here $m_{xxI}$ and $m_{IzI}$ are spin independent terms, while $m_{zzz}$, $m_{yyz}$ and $m_{zIz}$ originate from SOC. An example of moir\'e bands with type X substrate is shown in \cref{fig: information of all ideal flat bands part2}(d), which is calculated for substrate potential parameters (see the text in the figure) on the order of $\SI{10}{\milli\electronvolt}$ at $\theta=1.05^\circ$. The spin Chern numbers of the $|n|\le 2$ bands are $\mathcal{C}=\{-1, -1, 0, +2\}$ from high to low energies, and the quantum metric of the $n=+2$ band is nearly ideal (\cref{fig: information of all ideal flat bands part2}(e)). We leave the search for candidate type X substrate materials for future work.

\emph{Discussion}.---The commensurate substrate-induced intervalley coupling modifies TBG into an effective ``$\Gamma$-valley'' moir\'e model. The two TBG flat bands per valley become an effective $p_x$-$p_y$ two-orbital honeycomb lattice model with flat bands arising from geometric frustration, and spin Chern bands further emerge if the substrate has SOC. Among the type Y substrate candidates we identified in \cref{table: candidate substrate sqrt3 concise}, Sb$_2$Te$_3$ has a near-perfect lattice constant ratio $r_s \approx \sqrt{3}$ and is the most promising. A monolayer or at most few-layer ($\le 5$) Sb$_2$Te$_3$ substrate is desired to gap out its TI surface states \cite{Jiang_2012, kou2015}. It would be interesting if substrates without SOC can be found in the future, which would realize the pristine lattice model with geometric frustration in \cref{fig: information of all ideal flat bands part1}(b).

An interesting future direction is to explore how the insulating and superconducting phases in TBG \cite{wufc2018,Lian_2019,liucx2024,wangyj_2024} are modified by substrate-induced intervalley coupling. The spontaneous intervalley coherent (IVC) orders of the KIVC \cite{TBG4,Bultinck_2020} and TIVC states \cite{kwan2024TIVC,Nuckolls_2023} in TBG differ in momentum from the substrate-induced intervalley coupling here by $2\bold{q}_1$, so they may be significantly altered. The momentum of IVC order of the incommensurate Kekul\'e spiral (IKS) states \cite{Kwan_2021,Nuckolls_2023} may also be pinned by the substrate coupling. The high spin Chern number up to $4$ (which lower bounds the quantum metric) of flat bands with Sb$_2$Te$_3$ substrate may enhance the superfluid weight and temperature of superconductivity \cite{Xie_2020_prl_topology,yu2022,yu2025_Wilson}, and may favor chiral superconductivity \cite{han2025,geier2024,may-mann2025} or FCIs \cite{Ledwith2022prl,Ledwith2023prb,WangJie2021prl}. Lastly, it would be interesting to explore the possibility of spin liquids due to the lattice frustration nature of the effective model in \cref{eq:two-orbital-honeycomb} \cite{balents2010}.

\emph{Acknowledgments.} We thank Yong Xu, Hao-Wei Chen, Zijia Cheng, Jonah Herzog-Arbeitman, Minxuan Wang, Binghai Yan, Shuolong Yang and Yunhe Bai for helpful discussions. This work is supported by the National Science Foundation through Princeton University’s Materials Research Science and Engineering Center DMR-2011750, and the National Science Foundation under award DMR-2141966. Additional support is provided by the Gordon and Betty Moore Foundation through Grant GBMF8685 towards the Princeton theory program.

\bibliography{bibliography.bib}

\clearpage
\onecolumngrid

\setcounter{equation}{0}
\setcounter{figure}{0}
\setcounter{table}{0}
\setcounter{page}{1}
\makeatletter
\renewcommand{\theequation}{S\arabic{equation}}
\renewcommand{\thefigure}{S\arabic{figure}}
\renewcommand{\thetable}{S\arabic{table}}

\begin{center}
{\bf \large Supplemental Material for ``Intervalley-coupled Twisted Bilayer Graphene from  Substrate Commensuration''}
\end{center}

\section{Commensurate substrates}\label{appendix: Commensurate substrate}
We consider a system consisting of a graphene monolayer stacked on top of a substrate. In the context of the main text, this graphene monolayer is the lower layer of a twisted bilayer graphene (TBG) system, but we do not need to consider the top TBG layer in this section. Both the graphene and substrate layers are crystals with triangular Bravais lattices which we denote by $L_G$ and $L_s$, respectively. We say that $L_G$ and $L_s$ are commensurate if their intersection $L_{c} = L_{G} \cap L_{s}$ is not $\{\mathbf{0}\}$, and in this case $L_c$ is another triangular Bravais lattice \cite{scheer2023}. The ratio of the substrate's lattice constant to that of graphene is denoted by $r_s$, and the twist angle of the substrate relative to the graphene layer is denoted by $\phi_s$. Since most substrates have larger lattice constants than graphene, we assume $r_s \geq 1$. Additionally, we assume without loss of generality that $0 \le \phi_{s} \le \pi/6$. We aim to select only substrate configurations that couple $\boldK$ to $-\boldK$ in the graphene layer by requiring $\boldK$ and $-\boldK$ to be equivalent modulo $P_{c}$, the reciprocal lattice of $L_{c}$. In Appendix C5 of Ref. \cite{scheer2023}, this type of configuration was classified and denoted II+. Defining $\epsilon = \log r_s \geq 0$, $L_G$ and $L_s$ are commensurate of type II+ if and only if there exist integers $\mu, \nu, \and \rho$ satisfying $\mu\ge -3\nu\ge 0$, $\rho\ge \sqrt{\mu^2 + 3\nu^2}$, and
\begin{equation}
\label{eqn: exponential of epsilon and phi}
e^{-(\epsilon+i\phi_s)}=\frac{\mu+i\sqrt{3}\nu}{\rho}, \quad \text{gcd}(\mu, \nu, \rho)=1, \quad 3 \mid \rho.
\end{equation}

In this paper, we consider only substrates that preserve at least one of the mirror symmetries $M_{\hat{x}}$ and $M_{\hat{y}}$. This implies that $\phi_s = 0$ or $\phi_s = \pi/6$, and we now solve \cref{eqn: exponential of epsilon and phi} in these two cases.
\begin{enumerate}
    \item If $\phi_s=0$, \cref{eqn: exponential of epsilon and phi} implies $\nu=0$.
    Defining $\bar\rho = \rho/3$, the solutions of \cref{eqn: exponential of epsilon and phi} are $\text{gcd}(\mu, \bar\rho)=1$, $3\nmid \mu$, and $r_s = e^{\epsilon} = \frac{3\bar\rho}{\mu}$. Examples of $r_s$ include $\frac{3}{1}, \frac{3}{2}, \frac{6}{1}, \frac{6}{5}, \frac{9}{1}, \frac{9}{2}, \frac{9}{4}, \frac{9}{5}, \frac{9}{7}, \frac{9}{8}, \cdots$.
    
    \item If $\phi_s=\pi/6$, \cref{eqn: exponential of epsilon and phi} implies $\frac{\sqrt{3}\nu}{\mu} = \tan(-\phi_s)=-\frac{1}{\sqrt{3}}$, so that $\mu=-3\nu$. The imaginary part of \cref{eqn: exponential of epsilon and phi} becomes $e^{\epsilon} = -\frac{\rho}{2\sqrt{3} \nu}$. Defining $\bar\rho = \rho/3$, the solutions of \cref{eqn: exponential of epsilon and phi} are $\text{gcd}(\nu, \bar\rho)=1$, $3 \nmid \nu$, and $r_s = e^{\epsilon} = -\frac{\sqrt{3}\bar{\rho}}{2\nu}$. Examples of $r_s$ include $\sqrt{3}, \frac{3\sqrt{3}}{2}, \frac{3\sqrt{3}}{4}, 2\sqrt{3}, \frac{5\sqrt{3}}{2}, \frac{5\sqrt{3}}{4}, \frac{5\sqrt{3}}{8}, 3\sqrt{3}, \frac{3\sqrt{3}}{5}, \cdots$.
\end{enumerate}

We now consider the special case in which the substrate has a honeycomb lattice structure consisting of two triangular sublattices. If the sublattices of the substrate are of the same type (e.g., the material has only one atomic species) then the substrate is maximally symmetric and the system has symmetries $C_{3z}$, $C_{2z}$, $M_{\hat{x}}$, $M_{\hat{y}}$, and $\mathcal{T}$. If the two sublattices are inequivalent, $C_{2z}$ symmetry is broken, and only one of the mirror symmetries $M_{\hat{x}}$ and $M_{\hat{y}}$ remains. In the case of $\phi_s=0$, $M_{\hat{x}}$ is preserved, while for $\phi_s=\pi/6$, $M_{\hat{y}}$ is preserved.

Defining coprime integers $\xi_n$ and $\xi_d$ such that $r_s^2 = \frac{\xi_{n}}{\xi_{d}}$, the type II+ configurations with $\xi_{n}\xi_{d} \le 7$ are listed in \cref{table: Examples of commensurate substrates}. The third row (``symmetry'') in the table indicates the additional symmetry (other than $C_{3z}$ and $\mathcal{T}$) that the system retains when the two sublattices of the substrate are of different types. In Fig. 1 of the main text, we illustrate the configurations $(r_s, \phi_s)=\left(\frac{3}{2}, 0^{\circ}\right)$ and $(r_s, \phi_s)=(\sqrt{3}, 30^{\circ})$.

\begin{table}[h]
\centering
\begin{tabular}{||c||c|c|c|c|c|c|c|c|c||}
\hline
$r_s$ & $\frac{3}{2}$ & $\sqrt{3}$ & $3$ & $2\sqrt{3}$ & $\sqrt{21}$ & $3\sqrt{3}$ & $6$ & $\sqrt{39}$ & $4\sqrt{3}$ \\
\hline
$\phi_s$ & $0^\circ$ & $30^\circ$ & $0^\circ$ & $30^\circ$ & $10.89^\circ$ & $30^\circ$ & $0^\circ$ & $16.10^\circ$ & $30^\circ$ \\
\hline
symmetry & $M_{\hat{x}}$ & $M_{\hat{y}}$ & $M_{\hat{x}}$ & $M_{\hat{y}}$ & None & $M_{\hat{y}}$ & $M_{\hat{x}}$ & None & $M_{\hat{y}}$ \\
\hline
\end{tabular}
\caption{Examples of commensurate substrates with triangular Bravais lattices. The angles $\phi_s$ are given in degrees to two decimal places.}
\label{table: Examples of commensurate substrates}
\end{table}

\section{Hamiltonian from symmetry}
\label{appendix: Hamiltonian from symmetry}
In this section, we analyze single layer graphene, with and without substrate and spin-orbit coupling (SOC). We will show the most general form that a Hamiltonian can take that respects certain symmetries, such as $C_{3z}$, $C_{2z}$, $M_{\hat{x}}$, $M_{\hat{y}}$, and $\mathcal{T}$.

The Fermi level of a single graphene layer is located at momenta $\eta \boldK$ where $\eta=\pm$ is the valley index. Assuming small effects from the substrate distortion and SOC, we can perturbatively analyze the kinematics around these points. Consequently, we will examine the symmetries at the $\boldK$ point and use these symmetry constraints to determine the Hamiltonian.

In this section, we use $\tau_\mu$, $\sigma_\mu$, $s_\mu$ to denote the $2\times2$ identity ($\mu=0$) and Pauli ($\mu=x,y,z$) matrices in the basis of valley $\eta=\pm$, sublattice $\alpha = \pm$ and spin $s = \pm$, respectively. The projections onto each component are denoted as
\begin{equation}
\tau_{\pm} = \frac{\tau_0 \pm \tau_z}{2}
\,, \,\,
\sigma_{\pm} = \frac{\sigma_0 \pm \sigma_z}{2}
\,, \,\,
s_{\pm} = \frac{s_0 \pm s_z}{2}
\end{equation}
We define the vectors of Pauli matrices $\boldtau = (\tau_{x}, \tau_{y})$, $\boldsigma = (\sigma_{x}, \sigma_{y})$, $\bolds = (s_{x}, s_{y})$. Additionally, we use $R_\theta$ to denote rotation by angle $\theta$ about the $z$ axis, $\mathcal{R}_x$ to denote reflection across the $y$ axis (flipping $x$ to $-x$), and $\mathcal{R}_y$ to denote reflection across the $x$ axis  (flipping $y$ to $-y$).

\subsection{Without substrate and without SOC}
\begin{figure}[h]
\centering
\includegraphics[width=60mm]{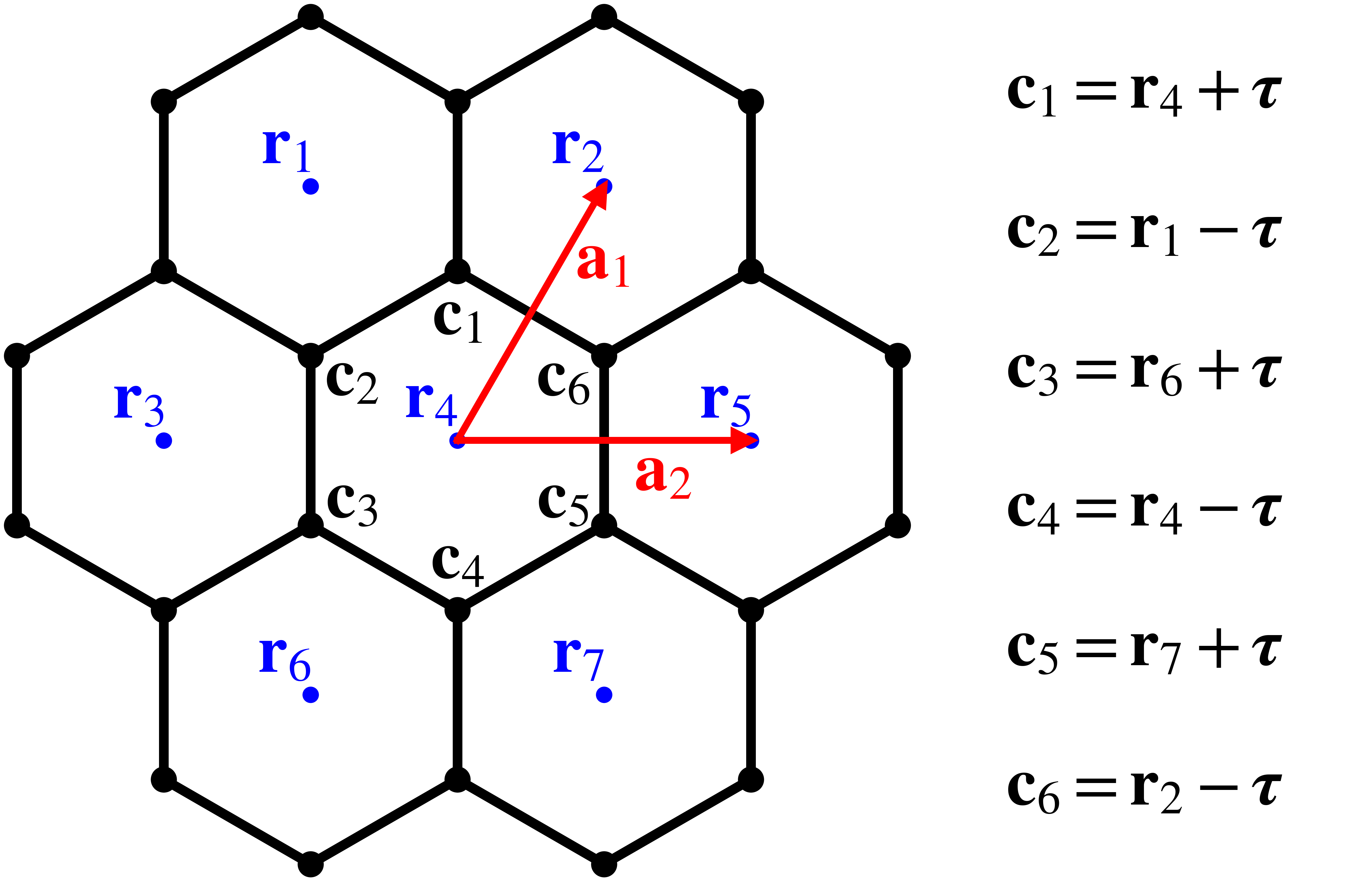}
\caption{Indexing atoms in graphene. The positions of the carbon atoms are denoted by $\mathbf{c}_{i}$, which are defined by the centers of the hexagons $\mathbf{r}_i$ and the sublattice indices $\alpha_i=\pm$ by $\mathbf{c}_{i} = \mathbf{r}_i + \alpha_i \boldsymbol{\tau}$.}
\label{subfig: graphene r alpha definition}
\end{figure}

We first study the Hamiltonian for a graphene layer without a substrate and without spin-orbit coupling (SOC). A single layer of graphene is formed from a 2D triangular lattice, with each unit cell containing two atoms. We denote the Bravais lattice consisting of the centers of the hexagons by $L$ (blue dots in \cref{subfig: graphene r alpha definition}), and the atomic positions can be represented by $\boldr + \alpha \boldtau$ for $\boldr \in L$ (black dots in \cref{subfig: graphene r alpha definition}). Denoting the orbital at site $\boldr+\alpha\boldtau$ by $\ket{\boldr, \alpha}$, the Bloch states are defined as
\begin{equation}\label{eqn: definition of the Bloch states}
\ket{\boldk, \alpha} = \frac{1}{\sqrt{\BZ}} \sum_{\boldr \in L} e^{i \boldk \cdot (\boldr+\alpha \boldtau)} \ket{\boldr, \alpha}
\end{equation}
where $|\text{BZ}|$ is the area of the Brillouin zone. Since we are assuming in this section that there is no SOC, we neglect the spin degrees of freedom. We expand the Bloch states around $\eta\boldK$ as $\ket{\boldp, \eta, \alpha} = \ket{\eta\boldK+\boldp, \alpha}$, and the symmetry operators act on these states as follows:
\begin{equation}
\begin{split}
C_{3z} \ket{\boldp, \eta, \alpha} &= e^{2 \pi i \alpha \eta / 3} \ket{R_{2\pi/3} \boldp, \eta, \alpha}
\\
C_{2z} \ket{\boldp, \eta, \alpha} &= \ket{-\boldp, -\eta, -\alpha}
\\
\mathcal{T} \ket{\boldp, \eta, \alpha} &= \ket{-\boldp, -\eta, \alpha}
\\
M_{\hat{y}} \ket{\boldp, \eta, \alpha} &= \ket{\mathcal{R}_{\hat{y}}\boldp, \eta, -\alpha}.
\end{split}
\end{equation}
We now focus on the $\eta = +$ valley. We represent the Hamiltonian in this valley by the matrix $H(\boldp)$ which satisfies 
\begin{equation}
\braket{\boldp', +, \alpha' | \hat{H} | \boldp, +, \alpha} = H(\boldp)_{\alpha',\alpha}\delta^2(\boldp' - \boldp).
\end{equation}
The symmetries $C_{3z}$, $C_{2z} \mathcal{T}$, and $M_{\hat{y}}$ constrain $H(\boldp)$ as follows:
\begin{equation}\label{eqn: constraints for single layer graphene without substrate and without SOC}
\begin{split}
C_{3z}^{-1} \hat{H} C_{3z} = \hat{H} 
&\Rightarrow e^{-2 \pi i \sigma_{z} / 3} H(R_{2\pi/3}\boldp) e^{2 \pi i \sigma_{z} / 3} = H(\boldp)
\\
(C_{2z} \mathcal{T})^{-1} \hat{H} (C_{2z} \mathcal{T}) =  \hat{H} 
&\Rightarrow \sigma_{x} H^{*}(\boldp) \sigma_{x} = H(\boldp)
\\
M_{\hat{y}}^{-1} \hat{H} M_{\hat{y}} = \hat{H} 
&\Rightarrow \sigma_{x} H(\mathcal{R}_{\hat{y}} \boldp) \sigma_{x} = H(\boldp).
\end{split}
\end{equation}

To first order in $\boldp$, the most general ansatz for $H(\boldp)$ takes the form
\begin{equation}
\begin{split}
H(\boldp) = \sum_{\mu=0}^{3} (c_{0 \mu} + c_{x  \mu} p_{x} + c_{y \mu} p_{y})\sigma_{\mu}.
\end{split}
\end{equation}
The conditions in \cref{eqn: constraints for single layer graphene without substrate and without SOC} fix the Hamiltonian to be
\begin{equation}
H(\boldp) = \hbar v_F  \boldp \cdot \boldsigma + E_F \sigma_{0},
\end{equation}
where $v_F$ is Fermi velocity and $E_F$ is the Fermi energy.

\subsection{With substrate and without SOC}
We now add a substrate layer which couples the $\pm \boldK$ points as described in \cref{appendix: Commensurate substrate}, but we continue to assume there is no SOC. We now define the matrix $H(\boldp)$ by
\begin{equation}
\braket{\boldp', \eta', \alpha' | \hat{H} | \boldp, \eta, \alpha} = H(\boldp)_{\eta'\alpha',\eta\alpha}\delta^2(\boldp' - \boldp),
\end{equation}
so that $H(\boldp)$ represents the Hamiltonian in both valleys. A maximally symmetric substrate satisfies the following symmetry constraints:
\begin{equation}\label{eqn: constraints for single layer graphene with substrate and without SOC}
\begin{split}
C_{3z}^{-1} \hat{H} C_{3z} = \hat{H} 
&\Rightarrow e^{-2 \pi i \tau_{z} \sigma_{z} / 3} H(R_{2\pi/3}\boldp) e^{2 \pi i \tau_{z} \sigma_{z} / 3} = H(\boldp)
\\
C_{2z}^{-1} \hat{H} C_{2z} = \hat{H} 
&\Rightarrow (\tau_{x} \sigma_{x})  H(-\boldp) (\tau_{x} \sigma_{x}) = H(\boldp)
\\
\mathcal{T}^{-1} \hat{H} \mathcal{T} = \hat{H} 
&\Rightarrow (\tau_{x} \sigma_{0}) H^{*}(-\boldp) (\tau_{x} \sigma_{0}) = H(\boldp)
\\
M_{\hat{x}}^{-1} \hat{H} M_{\hat{x}} = \hat{H}
&\Rightarrow (\tau_{x} \sigma_{0}) H(\mathcal{R}_{\hat{x}} \boldp) (\tau_{x} \sigma_{0}) = H(\boldp)
\\
M_{\hat{y}}^{-1} \hat{H} M_{\hat{y}} = \hat{H}
&\Rightarrow (\tau_{0} \sigma_{x}) H(\mathcal{R}_{\hat{y}} \boldp) (\tau_{0} \sigma_{x}) = H(\boldp),
\end{split}
\end{equation}
where $C_{2z} = M_{\hat{x}} M_{\hat{y}}$ and $R_{\theta}\boldp = (p_x \cos\theta - p_y\sin\theta, p_x\sin\theta + p_y\cos\theta)$. To first order in momentum $\boldp$, these symmetry constraints imply that the Hamiltonian takes the form
\begin{equation}
H(\boldp) = \hbar v_F \tau_{+} \left( \boldp \cdot \boldsigma \right)
- \hbar v_F \tau_{-} \left( \boldp \cdot \boldsigma^{*} \right)
+ m_{xxI} \tau_x \sigma_x + E_F \tau_0 \sigma_0.
\end{equation}
The presence of the substrate introduces a mass term $m_{xxI}$ that couples the two valleys. This Hamiltonian is studied in Ref. \cite{scheer2023prl}, in which lithium adatoms provide a Kekul\'e-O distortion, i.e., $(r_{s}, \phi_{s})=(\sqrt{3}, 30^{\circ})$.

\subsection{With substrate and with SOC}
For substrates with SOC, we need to include a spin index $s=\pm1$ which represents the $z$ component of spin ($\uparrow$ and $\downarrow$). The states are now denoted $\ket{\boldp, \eta, \alpha, s}$. The symmetry operators acts on these states as
\begin{equation}\label{eqn: symmetry operators with substrate and SOC}
\begin{split}
C_{3z} \ket{\boldp, \eta, \alpha, s} 
&= e^{2 \pi i \alpha \eta / 3} e^{-i\pi s/3} \ket{R_{2\pi/3} \boldp, \eta, \alpha, s}
\\
C_{2z} \ket{\boldp, \eta, \alpha, s} 
&= e^{-i\pi s/2} \ket{-\boldp, -\eta, -\alpha, s}
\\
\mathcal{T} \ket{\boldp, \eta, \alpha, s} 
&= s \ket{-\boldp, -\eta, \alpha, -s}
\\
M_{\hat{x}} \ket{\boldp, \alpha, s} 
&= -i \ket{\mathcal{R}_{\hat{x}}\boldp, -\eta, \alpha, -s}\\
M_{\hat{y}} \ket{\boldp, \alpha, s} 
&= s \ket{\mathcal{R}_{\hat{y}}\boldp, \eta, -\alpha, -s}.
\end{split}
\end{equation}
A Hamiltonian that respects all symmetries satisfies
\begin{equation}
\begin{split}
C_{3z}^{-1} \hat{H} C_{3z} = \hat{H} 
&\Rightarrow \left( e^{-2 \pi i \tau_{z} \sigma_{z} / 3} e^{i\pi s_{z}/3} \right) H(R_{2\pi/3}\boldp) 
\left(e^{2 \pi i \tau_{z} \sigma_{z} / 3} e^{-i\pi s_{z}/3} \right) = H(\boldp)
\\
C_{2z}^{-1} \hat{H} C_{2z} = \hat{H} 
&\Rightarrow \left(\tau_{x} \sigma_{x} e^{i\pi s_{z}/2} \right) H(-\boldp) 
\left( \tau_{x} \sigma_{x} e^{-i\pi s_{z}/2} \right) = H(\boldp)
\\
\mathcal{T}^{-1} \hat{H} \mathcal{T} = \hat{H} 
&\Rightarrow \left(i \tau_{x} \sigma_{0} s_{y} \right) H^{*}(-\boldp) 
\left(-i \tau_{x} \sigma_{0} s_{y} \right) = H(\boldp)
\\
M_{\hat{x}}^{-1} \hat{H} M_{\hat{x}} = \hat{H}
&\Rightarrow \left( \tau_{x} \sigma_{0} e^{i\pi s_{x}/2} \right) H(\mathcal{R}_{\hat{x}} \boldp) 
\left( \tau_{x} \sigma_{0} e^{-i\pi s_{x}/2} \right)= H(\boldp)
\\
M_{\hat{y}}^{-1} \hat{H} M_{\hat{y}} = \hat{H}
&\Rightarrow \left( \tau_{0} \sigma_{x} e^{i\pi s_{y}/2} \right) H(\mathcal{R}_{\hat{y}} \boldp) 
\left( \tau_{0} \sigma_{x} e^{-i\pi s_{y}/2} \right) = H(\boldp)
\end{split}
\end{equation}
and can be written to first order in $\boldp$ in the form
\begin{equation}\label{eqn: the most general Hamiltonian with SOC}
H = \sum_{i=1}^{10} c_{i} h_{i},
\end{equation}
where
\begin{equation}\label{eqn: eqn: terms of maximally symmetric substrate}
\begin{split}
h_0 & = \tau_{0} \sigma_{0} s_{0}
\\
h_1 &= \tau_+(\boldp \cdot \boldsigma) s_0 - \tau_- (\boldp \cdot \boldsigma^*) s_0
\\
h_2 &= \tau_x \sigma_x s_0
\\
h_3 &= \tau_z \sigma_z s_z
\\
h_4 &= \tau_y \sigma_y s_z
\\
h_5 &= \tau_0 \sigma_0 \left(\boldp \times \bolds \right)_{z}
\\
h_6 &= 
\tau_{+} \left( \boldsigma \times \bolds \right)_{z}
+ \tau_{-} \left( \boldsigma \times \bolds^{*} \right)_{z}
\\
h_7 &= \tau_x \sigma_x \left( \boldp \times \bolds \right)_{z}
\\
h_8 &=
\left[
\left( \boldp \cdot \boldtau \right) \sigma_{+}
- \left(\boldp \cdot \boldtau^{*} \right) \sigma_{-}
\right]
s_{z}
\\
h_9 &=
\tau_{0} \sigma_{x}
\left( \boldp \times \bolds^{*} \right)_{z}
-\tau_{z} \sigma_{y} 
\left( \boldp \cdot \bolds^{*} \right)
\\
h_{10} &=
\tau_{x} \sigma_{0} 
\left( \boldp \times \bolds^{*} \right)_{z}
-\tau_{y} \sigma_{z} 
\left( \boldp \cdot \bolds^{*} \right),
\end{split}
\end{equation}
where $\left(\boldp \times \bolds \right)_{z}$ denotes $\left(\boldp \times \bolds \right)\cdot\hat{\textbf{z}}$. Here, $h_0$ is a constant term, $h_1$ is the kinetic term, $h_2$, $h_3$, $h_4$ are momentum and spin independent, and $h_5$ represents Rashba SOC.

For a Hamiltonian that contains only the spin $s_z$ conserving terms $h_1, h_2, h_3, h_4$, we can express it in terms of a single spin component, such as spin-up, and discuss the resulting ``spinless Hamiltonian'':
\begin{equation}
H_{\uparrow}(\boldp) = \hbar v_F \tau_+(\boldp \cdot \boldsigma) - \hbar v_F \tau_-(\boldp \cdot \boldsigma^*)
+ m_{xxI} \tau_x \sigma_x
+ m_{yyz} \tau_y \sigma_y
+ m_{zzz} \tau_z \sigma_z
,\end{equation}
where we renamed the coefficients $(c_1, c_2, c_3, c_4) \rightarrow (\hbar v_F, m_{xxI}, m_{zzz}, m_{yyz})$. Here, we did not include the constant term $h_0$ since its only effect is a constant energy shift.

If the two sublattices in the substrate are different, the system breaks $C_{2z}$ symmetry. For simplicity, we consider configurations where either $M_{\hat{x}}$ or $M_{\hat{y}}$ is preserved, as discussed in \cref{appendix: Commensurate substrate}. In both cases, 7 extra terms are allowed:
\begin{enumerate}
\item $C_{3z}+M_{\hat{x}}+\mathcal{T}$: type-X substrate
\begin{equation}
\begin{aligned}
\label{eqn: type-X substrate general form}
h_{1}^{x} &= \tau_{z} \sigma_{0} s_{z}
\\
h_{2}^{x} &= \tau_{0} \sigma_{z} s_{0}
\\
h_{3}^{x} &= \tau_0 \sigma_{z} 
\left( \boldp \times \bolds \right)_{z}
\\
h_{4}^{x} &= 
\tau_{y} \sigma_{x} 
\left( \boldp \cdot \bolds \right)
\\
h_{5}^{x} &= 
\tau_{+} 
\left( \boldp \cdot \boldsigma \right) s_{z}
+ \tau_{-}
\left( \boldp \cdot \boldsigma^{*} \right) 
s_{z}
\\
h_{6}^{x} &= 
\left( \boldp \cdot \boldtau \right) \sigma_{+} s_{z}
+ \left( \boldp \cdot \boldtau^{*} \right) \sigma_{-} s_{z}
\\
h_{7}^{x} &=
\tau_{x} \sigma_{z} 
\left( \boldp \times \bolds^{*} \right)_{z}
-\tau_{y} \sigma_{0} 
\left( \boldp \cdot \bolds^{*} \right)
\end{aligned}
\end{equation}

\item $C_{3z}+M_{\hat{y}}+\mathcal{T}$: type-Y substrate
\begin{equation}
\begin{aligned}
\label{eqn: type-Y substrate general form}
h_{1}^{y} &= \tau_{x} \sigma_{y} s_{z}
\\
h_{2}^{y} &= \tau_{y} \sigma_{x} s_{0}
\\
h_{3}^{y} &= \tau_0 \sigma_{z} \left( \boldp \cdot \bolds \right)
\\
h_{4}^{y} &= \tau_{y} \sigma_{x} \left( \boldp \times \bolds \right)_{z}
\\
h_{5}^{y} &= 
\tau_{+} \left( \boldp \times \boldsigma \right)_{z} s_{z}
+ \tau_{-} \left( \boldp \times \boldsigma^{*} \right)_{z} s_{z}
\\
h_{6}^{y} &= 
\left( \boldp \times \boldtau \right)_{z}
\sigma_{+} s_{z}
+ \left( \boldp \times \boldtau^{*} \right)_{z} 
\sigma_{-} s_{z}
\\
h_{7}^{y} &=
\tau_{x} \sigma_{z} 
\left( \boldp \cdot \bolds^{*} \right)
+\tau_{y} \sigma_{0} 
\left( \boldp \times \bolds^{*} \right)_{z}
\end{aligned}
\end{equation}
\end{enumerate}

In this paper, we will only focus on the momentum-independent and spin $s_z$ preserving terms $h_{1}^{x}$, $h_{2}^{x}$, $h_{1}^{y}$, $h_{2}^{y}$.

\section{Symmetries}\label{appendix: symmetries}
In this section, we discuss the symmetries of the Hamiltonian, and show that Hamiltonians with different mass terms may be unitarily related, and therefore possess the same spectrum. Notably, since we do not take the small-angle approximation (i.e., the approximation of $\boldsigma_{l\theta/2}$ by $\boldsigma$ in main text Eq. (2)), the Hamiltonian does not have particle-hole symmetry.

The Bistritzer-MacDonald (BM) model \cite{Bistritzer2011} is a low energy continuum model for twisted bilayer graphene. The BM model can be written in the form
\begin{equation}
H_{0}(\boldr) = 
\begin{pmatrix}
\mathcal{H}_{+}^{(0)}(\boldr) & \mathcal{H}_{\text{hop}}(\boldr) \\
\mathcal{H}_{\text{hop}}^{\dagger}(\boldr) & \mathcal{H}_{-}^{(0)}(\boldr)
\end{pmatrix}
\end{equation}
in the basis $\ket{\boldr, l, \eta, \alpha, s}$, where $l = +$ ($l = -$) denotes the top (bottom) layer. We use $\Gamma_\mu$ to denote the $2\times 2$ identity ($\mu=0$) and Pauli ($\mu=x, y, z$) matrices in the layer basis $l$, and define $\Gamma_{\pm} = \frac{1}{2} (\Gamma_0 \pm \Gamma_z)$. Adding a maximally symmetric substrate (i.e., one which has $C_{3z}$, $C_{2z}$, $M_{\hat{y}}$, and $\mathcal{T}$ symmetries) to the bottom layer introduces a distortion
\begin{equation}
\Delta H(\boldr) = 
\begin{pmatrix}
0 & 0 \\
0 & \mathcal{H}_{-}^{(\text{sub})}(\boldr)
\end{pmatrix}
.\end{equation}
In principle, $\mathcal{H}_{-}^{(\text{sub})}(\boldr)$ includes all terms from \cref{eqn: eqn: terms of maximally symmetric substrate} (for maximally symmetric, type-X, and type-Y substrates), in \cref{eqn: type-X substrate general form} (for type-X substrate), or in \cref{eqn: type-Y substrate general form} (for type-Y substrate). It was shown in \cite{scheer2023} that in the case of $r_s = \sqrt{3}$ and $\phi_s = 30^{\circ}$ the substrate induced SOC is $s_z$ conserving under reasonable approximations. For simplicity, we also assume here that the Hamiltonian preserves $s_z$. Additionally, we expect that momentum independent substrate potential terms have a greater effect than momentum dependent terms because the momenta relevant to the low energy physics include only small deviations from the $\boldK$ and $-\boldK$ points of graphene. We therefore only include momentum independent substrate potential terms. Specifically, the terms we keep are:
\begin{equation}
\begin{cases}
\label{eqn: all mass terms}
\text{maximally symmetric, type-X, and type-Y:}& 
\tau_x \sigma_x s_0, \,
\tau_z \sigma_z s_z, \,
\tau_y \sigma_y s_z
\\
\text{type-X:}&
\tau_{z} \sigma_{0} s_{z}, \,
\tau_{0} \sigma_{z} s_{0}
\\
\text{type-Y:}&
\tau_{x} \sigma_{y} s_{z}, \,
\tau_{y} \sigma_{x} s_{0}
\end{cases}
\end{equation}

Since we do not take the small-angle approximation, the sublattice potential must be modified by a rotation of $-\theta/2$, to account for the rotation of the bottom graphene layer. Therefore, the mass terms in \cref{eqn: all mass terms} should be unitarily transformed by the rotation operator
\begin{equation}
U = e^{-i\theta \tau_{z}\sigma_{z} / 4} e^{-i\theta s_{z}/4}
,\end{equation}
which transforms the $\eta = s = +$ Dirac kinetic term $\boldp\cdot\boldsigma$ to $\boldp\cdot\boldsigma_{-\theta/2}$. However, it is easy to see that all the mass terms in \cref{eqn: all mass terms} commute with $U$. Therefore their form is unchanged by this transformation.

In the following, we often work in moir\'e momentum space. The moir\'e crystal momentum $\boldk$ for a state $\ket{\psi}$ is defined by $T_\boldR \ket{\psi} = e^{-i\boldk \cdot \boldR} \ket{\psi}$ where $T_\boldR$ is the translation operator defined in the main text. Furthermore, $H(\boldk)$ denotes the representation of the Hamiltonian in a plane wave basis consistent with this definition of moir\'e crystal momentum.

\subsection{Maximally symmetric substrate with $C_{2z}$ symmetry}
\label{appendix: Symmetries of a maximally symmetric substrate}
When the substrate is maximally symmetric, $\mathcal{H}_{-}^{(\text{sub})}$ contains 3 terms:
\begin{equation}
\mathcal{H}_{-}^{(\text{sub})}
= m_{xxI} \tau_x \sigma_x s_0 
+ m_{zzz} \tau_z \sigma_z s_z 
+ m_{yyz} \tau_y \sigma_y s_z 
.\end{equation}
We denote the Hamiltonian with mass terms $m_{xxI}$, $m_{zzz}$, and $m_{yyz}$ as 
\begin{equation}
H(\boldr, m_{xxI}, m_{zzz}, m_{yyz}) 
= H_{0}(\boldr) + \Gamma_{-}\mathcal{H}_{-}^{(\text{sub})}
= H_{0}(\boldr) 
+ \Gamma_{-} \left( m_{xxI} \tau_x \sigma_x s_0 
+ m_{zzz} \tau_z \sigma_z s_z 
+ m_{yyz} \tau_y \sigma_y s_z \right).
\end{equation}
The Hamiltonian respects time-reversal symmetry $\mathcal{T}$, which acts on the state $\ket{\boldr, l, \eta, \alpha, s}$ as
\begin{equation}
\mathcal{T} \ket{\boldr, l, \eta, \alpha, s} = s \ket{\boldr, l, -\eta, \alpha, -s}.
\end{equation}
According to the Kramers' theorem, the spectrum is 2-fold degenerate at time-reversal invariant momenta, namely $\boldGamma_M$ and $R_{\zeta_j}\boldM_M$ for $j \in \{1,2,3\}$. Each degenerate state belongs to a Kramers pair, with opposite spin components.

Focusing on a particular spin component, when $m_{xxI}=m_{yyz}=0$ the two valleys are decoupled and the Hamiltonian decomposes as a direct sum $H(\boldk)=H_{\eta=+}(\boldk) \oplus H_{\eta=-}(\boldk)$. Under $C_{2z}$, the components transform into one another, $C_{2z}^{-1} H_{\eta}(\boldk) C_{2z} = H_{-\eta}(-\boldk)$, resulting in degeneracies at $\boldGamma_M$ and $\boldM_M$, as we now explain.
\begin{enumerate}
\item $\boldk=\boldGamma_M$: In this case,
\begin{equation}
C_{2z}^{-1} H_{\eta}(\boldGamma_M) C_{2z} = H_{-\eta}(-\boldGamma_M) = H_{-\eta}(\boldGamma_M)
,\end{equation}
therefore the two components of the direct sum are related by a unitary transformation $C_{2z}$, resulting in the degeneracy at $\boldGamma_M$.
\item $\boldk=\boldM_M$: $\boldM_M$ and $-\boldM_M$ are related by a reciprocal lattice translation, and momenta related by reciprocal lattice $\boldG$ can be transformed into one another by the unitary \emph{embedding matrix} $V(\boldG)$:
\begin{equation}
H(\boldk+\boldG) = V(\boldG) H(\boldk) V^{-1}(\boldG)
.\end{equation}
\noindent
As a result, the two components are related by
\begin{equation}
\begin{aligned}
C_{2z}^{-1} H_{\eta}(\boldM_M) C_{2z} 
&= H_{-\eta}(-\boldM_M) 
\\
&= V(\boldG) H_{-\eta}(\boldGamma_M) V^{-1}(\boldG)
\\
\Rightarrow H_{-\eta}(\boldM_M) 
&= V^{-1}(\boldG) C_{2z}^{-1} H_{-\eta}(\boldM_M) C_{2z} V(\boldG)
.\end{aligned}
\end{equation}
The two components are again related by a unitary transformation $C_{2z} V(\boldG)$, and the spectrum is degenerate at $\boldM_M$.
\end{enumerate}

While the three mass terms may seem independent, Hamiltonians with different masses are sometimes unitarily equivalent. To see this, we introduce two unitary (and hermitian) operators $U_{zII}$ and $U_{IIx}$ and compute their action on the Hamiltonian:
\begin{enumerate}
\item $U_{zII}$:
\begin{equation}
\begin{aligned}
\label{eqn: definition of UzII}
&U_{zII} \ket{\boldr, l, \eta, \alpha, s} = \eta \ket{\boldr, l, \eta, \alpha, s}
\\
&U_{zII}^{\dagger} H(\boldr, m_{xxI}, m_{zzz}, m_{yyz}) U_{zII}^{\dagger} = H(\boldr, -m_{xxI}, m_{zzz}, -m_{yyz})
.\end{aligned}
\end{equation}
\item $U_{IIx}$:
\begin{equation}
\begin{aligned}
\label{eqn: definition of UIIx}
&U_{IIx} \ket{\boldr, l, \eta, \alpha, s} = \ket{\boldr, l, -\eta, -\alpha, -s}
\\
&U_{IIx}^{\dagger} H(\boldr, m_{xxI}, m_{zzz}, m_{yyz}) U_{IIx}^{\dagger} = H(\boldr, m_{xxI}, -m_{zzz}, -m_{yyz})
.\end{aligned}
\end{equation}
\end{enumerate}
This implies that, instead of exploring the enitre phase space with arbitrary $m_{xxI}$, $m_{zzz}$, and $m_{yyz}$, we only need to focus on regions with $m_{xxI}>0$ and $m_{yyz}>0$; the rest of the phase space can be inferred from these results.

\subsection{Type-Y substrate}
For a type-Y substrate, where the $M_{\hat{x}}$ symmetry is broken, $\mathcal{H}_{-}^{(\text{sub})}$ admits more spin-conserving terms that are momentum independent, and the Hamiltonian is represented by (see \cref{eqn: type-Y substrate general form})
$$
H(\boldr, m_{xxI}, m_{zzz}, m_{yyz}, m_{xyz}, m_{yxI}) 
= H_{0} 
+ \Gamma_{-} \left( m_{xxI} \tau_x \sigma_x s_0 
+ m_{zzz} \tau_z \sigma_z s_z 
+ m_{yyz} \tau_y \sigma_y s_z 
+ m_{xyz} \tau_x \sigma_y s_z 
+ m_{yxI} \tau_y \sigma_x s_0 \right)
.$$
The five parameters can be reduced to four by observing that
\begin{equation}
\begin{aligned}
e^{i \chi \tau_z / 2} ( \tau_{x} \sigma_{x} s_{0}) e^{-i \chi \tau_z / 2}
&= \tau_{x} \sigma_{x} s_{0} \cos\chi - \tau_{y} \sigma_{x} s_{0} \sin\chi
\\
e^{i \chi \tau_z / 2} ( \tau_{y} \sigma_{x} s_{0}) e^{-i \chi \tau_z / 2}
&= \tau_{x} \sigma_{x} s_{0} \sin\chi + \tau_{y} \sigma_{x} s_{0} \cos\chi
\\
e^{i \chi \tau_z / 2} ( \tau_{x} \sigma_{y} s_{z}) e^{-i \chi \tau_z / 2}
&= \tau_{x} \sigma_{y} s_{z} \cos\chi - \tau_{y} \sigma_{y} s_{z} \sin\chi
\\
e^{i \chi \tau_z / 2} ( \tau_{y} \sigma_{y} s_{z}) e^{-i \chi \tau_z / 2}
&= \tau_{x} \sigma_{y} s_{z} \sin\chi + \tau_{y} \sigma_{y} s_{z} \cos\chi
,\end{aligned}
\end{equation}
which gives
\begin{equation}
\begin{aligned}
e^{i \chi \tau_z / 2} H(\boldr, m_{xxI}, m_{zzz}, m_{yyz}, m_{xyz}, m_{yxI}) e^{-i \chi \tau_z / 2}
= H(\boldr, m_{xxI}^{\prime}, m_{zzz}, m_{yyz}^{\prime}, m_{xyz}^{\prime}, m_{yxI}^{\prime})
,\end{aligned}
\end{equation}
where
\begin{equation}
\begin{aligned}
\begin{pmatrix}
m_{xxI}^{\prime} & m_{xyz}^{\prime} \\
m_{yxI}^{\prime} & m_{yyz}^{\prime}
\end{pmatrix}
&=
\begin{pmatrix}
\cos\chi & \sin\chi \\
-\sin\chi & \cos\chi
\end{pmatrix}
\begin{pmatrix}
m_{xxI} & m_{xyz} \\
m_{yxI} & m_{yyz}
\end{pmatrix}
.\end{aligned}
\end{equation}
This implies that $H(\boldr, m_{xxI}, m_{zzz}, m_{yyz}, m_{xyz}, m_{yxI})$ and $H(\boldr, m_{xxI}^{\prime}, m_{zzz}, m_{yyz}^{\prime}, m_{xyz}^{\prime})$ are unitarily related by $e^{-i \chi \tau_z / 2}$. Given any set of mass parameters, we can always choose $\chi$ such that $\tan\chi = m_{yxI}/m_{xxI}$. With this choice, we set $m_{yxI}^{\prime}=0$, reducing the number of independent mass terms to four effectively: : $m_{xxI}$, $m_{zzz}$, $m_{yyz}$, and $m_{xyz}$. Physically, this corresponds to a redefinition of the valleys in both layers. Therefore, the parameter $m_{yxI}$ is a redundant variable, and we omit it from further consideration.

In the presence of $m_{xyz}$, operators introduced in \cref{eqn: definition of UzII} and \cref{eqn: definition of UIIx} relate Hamiltonians with different mass terms:
\begin{enumerate}
\item $U_{zII}^{\dagger} H(\boldr, m_{xxI}, m_{zzz}, m_{yyz}, m_{xyz}) U_{zII} = H(\boldr, -m_{xxI}, m_{zzz}, -m_{yyz}, -m_{xyz})$.
\item $U_{IIx}^{\dagger} H(\boldr, m_{xxI}, m_{zzz}, m_{yyz}, m_{xyz}) U_{IIx} = H(\boldr, m_{xxI}, -m_{zzz}, -m_{yyz}, -m_{xyz})$.
\end{enumerate}
Due to the different sublattices, the $C_{2z}$ operator is no longer a symmetry; however, it relates Hamiltonians with different signs of $m_{xyz}$:
\begin{enumerate}[resume]
\item $C_{2z}$:
\begin{equation}
\begin{aligned}
&C_{2z} \ket{\boldr, l, \eta, \alpha, s} = \ket{-\boldr, l, -\eta, -\alpha, s}
\\
&C_{2z}^{\dagger} H(\boldr, m_{xxI}, m_{zzz}, m_{yyz}, m_{xyz}) C_{2z} = H(-\boldr, m_{xxI}, m_{zzz}, m_{yyz}, -m_{xyz})
.\end{aligned}
\end{equation}
This implies that the energy spectra of the two Hamiltonians are identical under a $\pi$-rotation around the $z$ axis, and the spin Chern numbers are the same for bands that differ in the spin component. 
\end{enumerate}
These relations allow us to complete the entire phase diagram of the band topology given the information in regions where $m_{xxI}>0$, $m_{yyz}>0$, and $m_{xyz}>0$.

\subsection{Type-X substrate}
A X-Type substrate with broken $M_{\hat{y}}$ symmetry admits spin-preserving and momentum independent terms in $\mathcal{H}_{-}^{(\text{sub})}$ such that the Hamiltonian takes the general form (see \cref{eqn: type-X substrate general form})
\begin{equation}
H(\boldr, m_{xxI}, m_{zzz}, m_{yyz}, m_{zIz}, m_{IzI}) 
= H_{0}(\boldr)
+ \Gamma_{-} \left( m_{xxI} \tau_x \sigma_x s_0 
+ m_{zzz} \tau_z \sigma_z s_z 
+ m_{yyz} \tau_y \sigma_y s_z 
+ m_{zIz} \tau_z \sigma_{0} s_z 
+ m_{IzI} \tau_{0} \sigma_z s_0 \right)
.\end{equation}
Similar to the previous cases, Hamiltonians with different mass terms are unitarily related:
\begin{enumerate}
\item $U_{zII}$:
\begin{equation}
\begin{aligned}
U_{zII}^{\dagger} H(\boldr, m_{xxI}, m_{zzz}, m_{yyz}, m_{zIz}, m_{IzI}) U_{zII} = H(\boldr, -m_{xxI}, m_{zzz}, -m_{yyz}, m_{zIz}, m_{IzI})
.\end{aligned}
\end{equation}
\item $U_{IIx}$:
\begin{equation}
\begin{aligned}
&U_{IIx}^{\dagger} H(\boldr, m_{xxI}, m_{zzz}, m_{yyz}, m_{zIz}, m_{IzI}) U_{IIx} = H(\boldr, m_{xxI}, -m_{zzz}, -m_{yyz}, -m_{zIz}, m_{IzI})
.\end{aligned}
\end{equation}
\item $C_{2z}$:
\begin{equation}
\begin{aligned}
C_{2z}^{\dagger} H(\boldr, m_{xxI}, m_{zzz}, m_{yyz}, m_{zIz}, m_{IzI}) C_{2z} = H(-\boldr, m_{xxI}, m_{zzz}, m_{yyz}, -m_{zIz}, -m_{IzI})
.\end{aligned}
\end{equation}
\end{enumerate}
By leveraging these relations, we can construct the full phase diagram of the band topology based on the information where $m_{xxI}>0$, $m_{yyz}>0$, and $m_{zIz}>0$.

\section{Fragile topology in the $m_{xxI}$ model}

\begin{figure}[h]
\centering
\includegraphics[width=140mm]{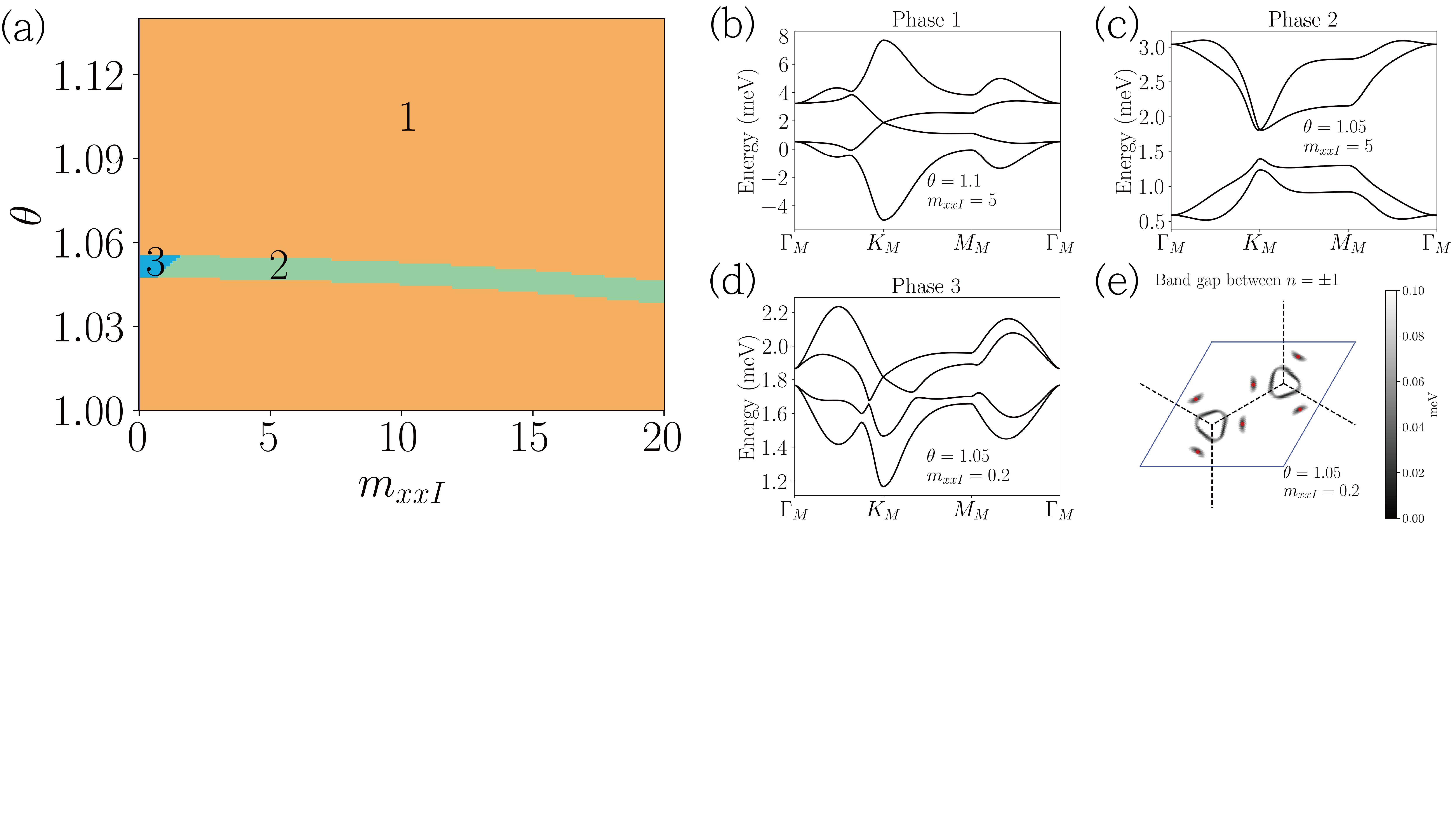}
\caption{(a) The phase diagram for the low-energy bands $n=-2\sim +2$. (b) In phase 1 (orange), four bands are connected at the $\Gamma_M$ and the $K_M$ points, forming an EBCR of $(\prescript{1}{}{E}\prescript{2}{}{E})_{2b}$. (c) In phase 2 (green), the top two and bottom two bands become gapped. The top two bands transform as the EBCR $(\prescript{1}{}{E}_{1}\prescript{2}{}{E}_{1})_{1a}$, while the bottom two transform as $(\prescript{1}{}{E}\prescript{2}{}{E})_{2b} \boxminus (\prescript{1}{}{E}_{1}\prescript{2}{}{E}_{1})_{1a}$. (d) In phase 3 (blue), the $n=\pm 1$ bands touch at non-high-symmetry points. Although the band structure appears as if the bands are separated, they transform as the EBCR $(\prescript{1}{}{E}\prescript{2}{}{E})_{2b}$ as a whole. (e) The band gap between $n=\pm 1$ for $\theta=1.05$ and $m_{xxI}=0.2$ across the entire BZ, with red dots indicating the band-touching points.}
\label{fig: TQC}
\end{figure}

Fragile topology refers to a type of band topology where a non-trivial set of bands can be trivialized by adding certain trivial bands. In contrast, stable topological bands can only be trivialized by combining with other non-trivial bands. For substrates that do not induce sufficiently strong SOC, the Hamiltonian contains only the $m_{xxI}$ term. In this case, the four low-energy bands ($n=-2\sim+2$) realize elementary band corepresentation (EBCR) of $(\prescript{1}{}{E}\prescript{2}{}{E})_{2b}$ of space group $P61^\prime$. A complete table of EBCRs for each magnetic space group is available on the Bilbao Crystallographic Server \cite{Bilbao, Bilbao_I, Bilbao_II, Elcoro_2021, Xu_magnetic}. In \cref{fig: TQC}(a), we show the phase diagram of the EBCR of the top two bands across various $\theta$ and $m_{xxI}$. Typical band structures for each phase are shown in \cref{fig: TQC}(b)-(d).
Interestingly, for certain values of the twisted angle $\theta$ and $m_{xxI}$ (phase 2 in \cref{fig: TQC}(a)), the top two bands ($n=+1, +2$) and the bottom two bands ($-2, -1$) become gapped (see \cref{fig: TQC}(c)), and the lower two bands exhibit fragile topology .

\section{Candidate Materials}

\begin{table}[t]
\centering
\begin{tabular}{|c|c|c|c|c|c|c|c|}
\hline
\textbf{substrate} & \textbf{$m_{xxI}$} &\textbf{$m_{zzz}$} & \textbf{$m_{yyz}$} &\textbf{$m_{xyz}$} & \textbf{$a_{s}$} & \textbf{$r_s / \sqrt{3}$} & \textbf{layer spacing} \\ \hline
$\text{Sb}_2 \text{Te}_3$  & 9.2 & 13.6 & 9.1 & 0.25 & 4.26 &  0.9998 & 3.498 \\ \hline
$\text{Ge} \text{Sb}_2 \text{Te}_4$  & 8.9 & 5.7 & 6.3 & 4.4 & 4.299 &  1.009 & 3.485 \\ \hline
\end{tabular}
\caption{Table of candidate substrates. All mass terms $m_{xxI}$, $m_{zzz}$, $m_{yyz}$, and $m_{xyz}$ are given in units of meV. The layer spacing, measured in $\AA$, represents the distance between the graphene and the uppermost atom of the substrate.}
\label{table: candidate substrate sqrt3}
\end{table}

\begin{figure}[t]
\centering
\subfigure[]{\includegraphics[width=40mm]{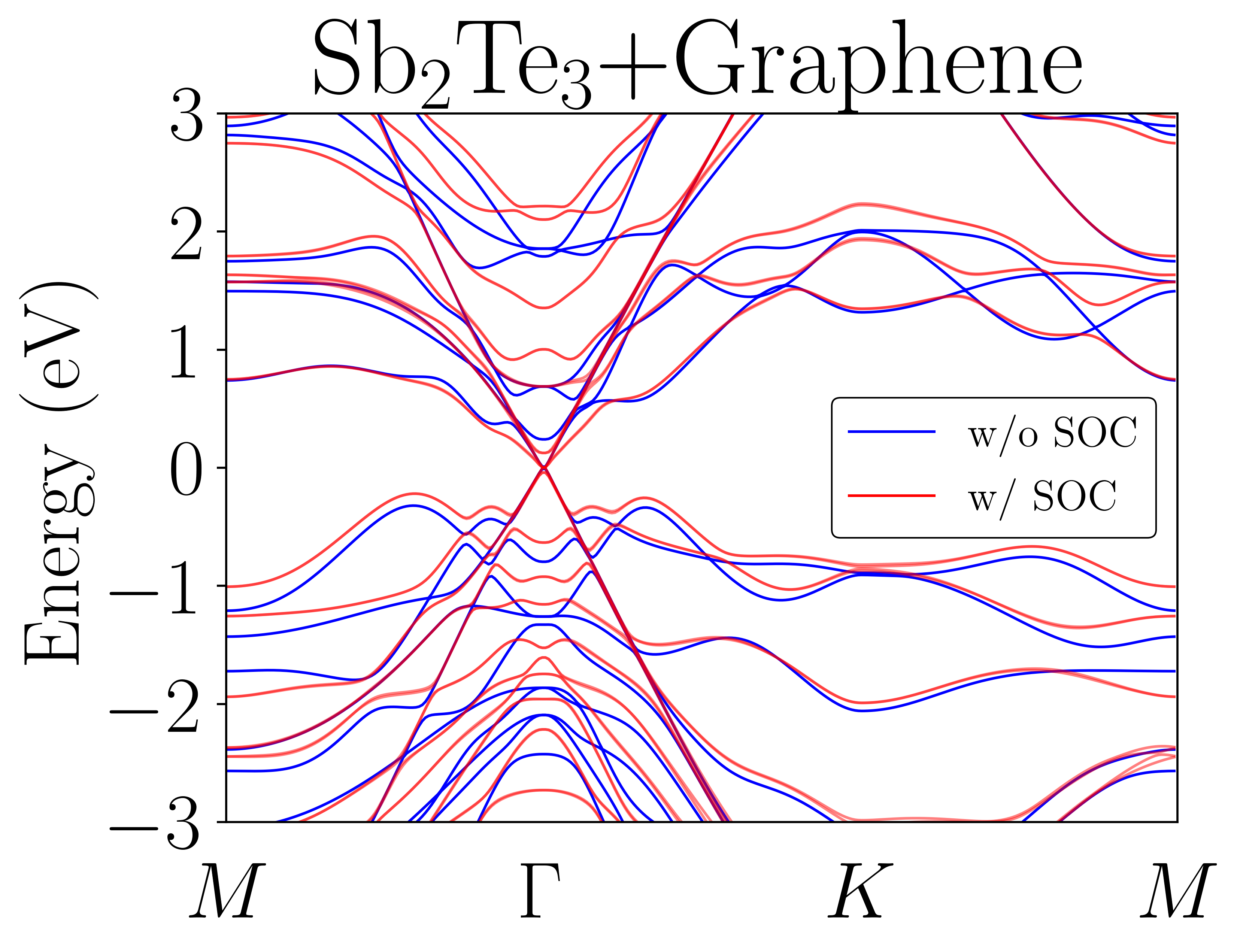}\label{subfig: band structure Sb2Te3Graphene}}
\subfigure[]{\includegraphics[width=40mm]{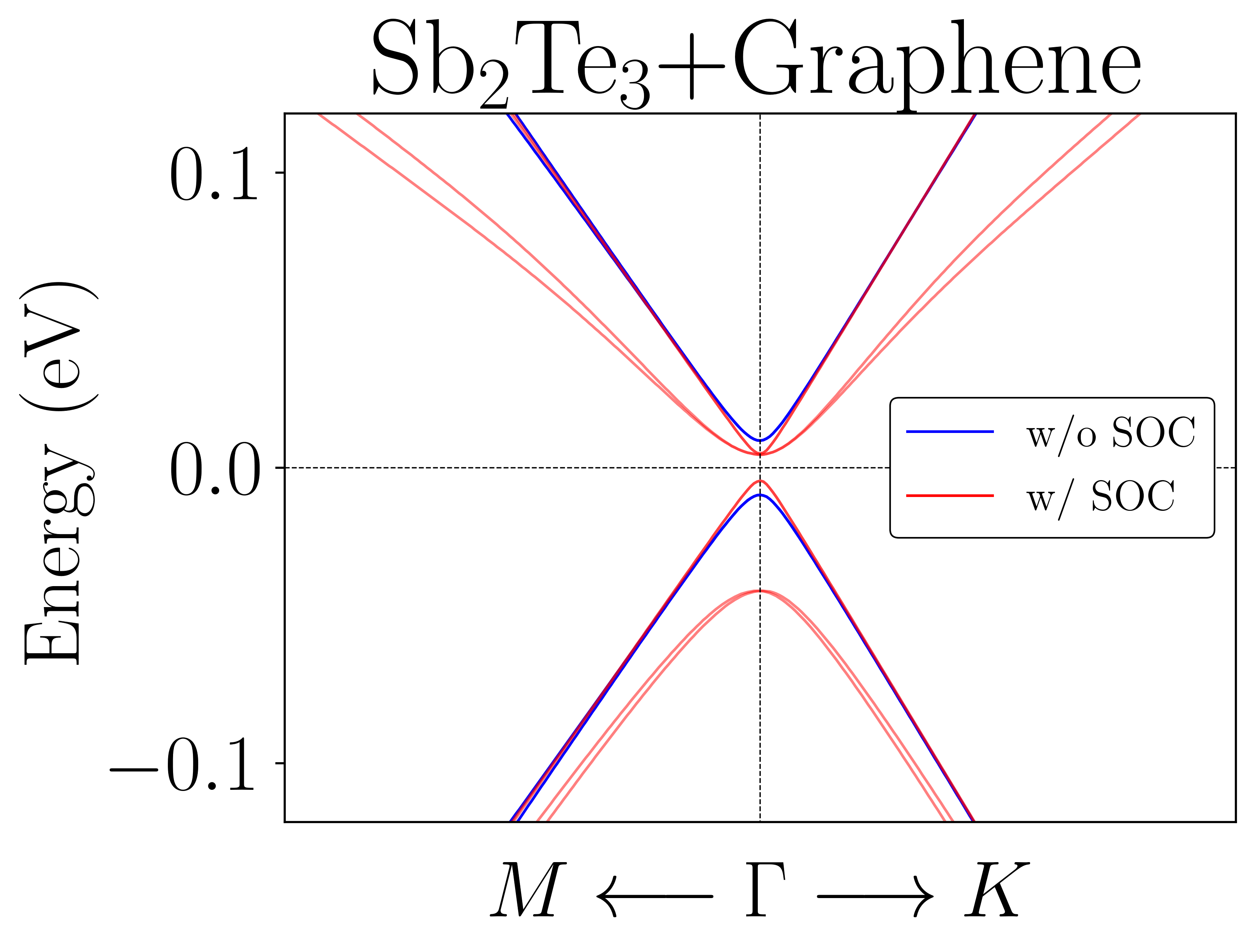}\label{subfig: band structure Sb2Te3Graphene zoom in}}
\subfigure[]{\includegraphics[width=40mm]{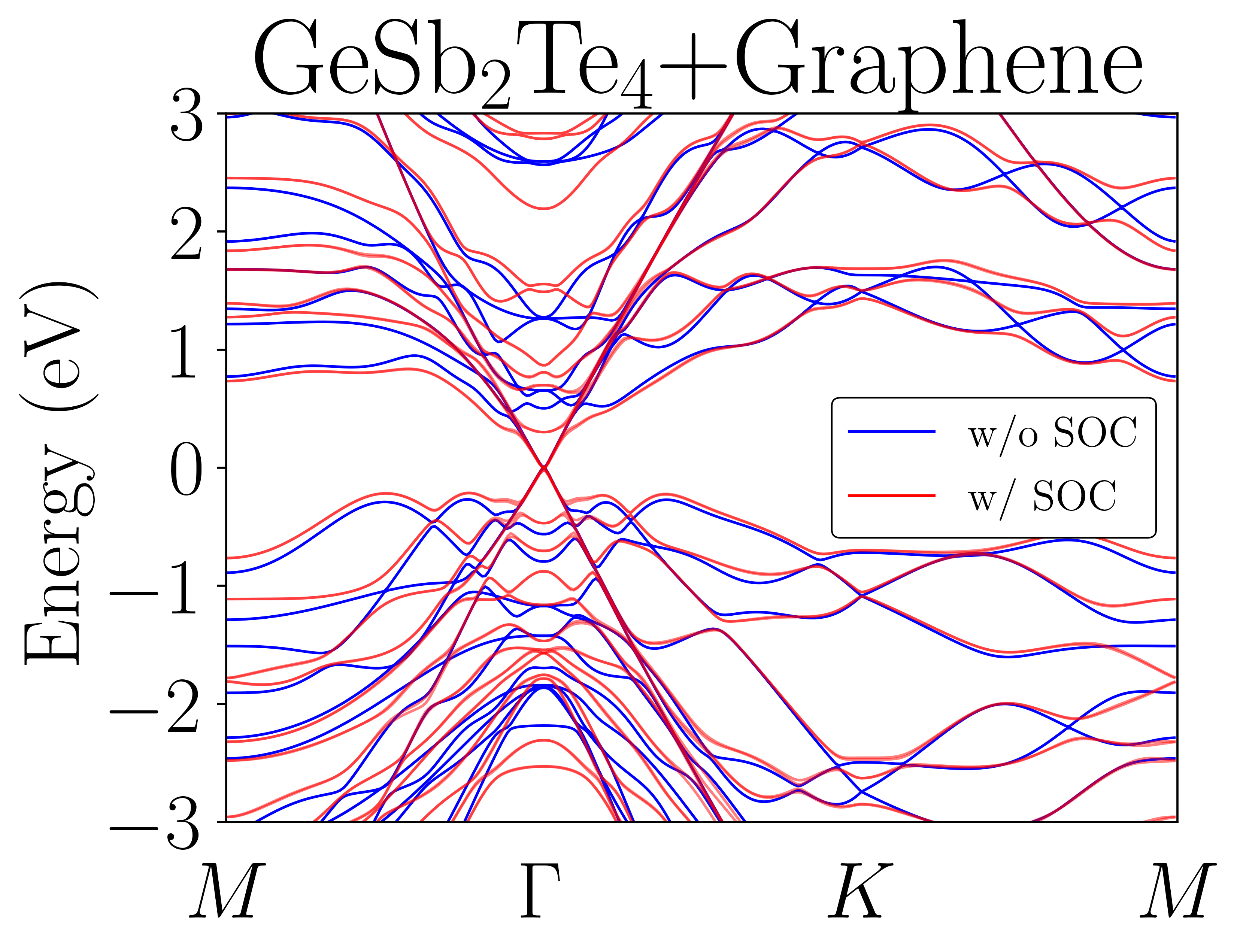}\label{subfig: band structure GeSb2Te4Graphene}}
\subfigure[]{\includegraphics[width=40mm]{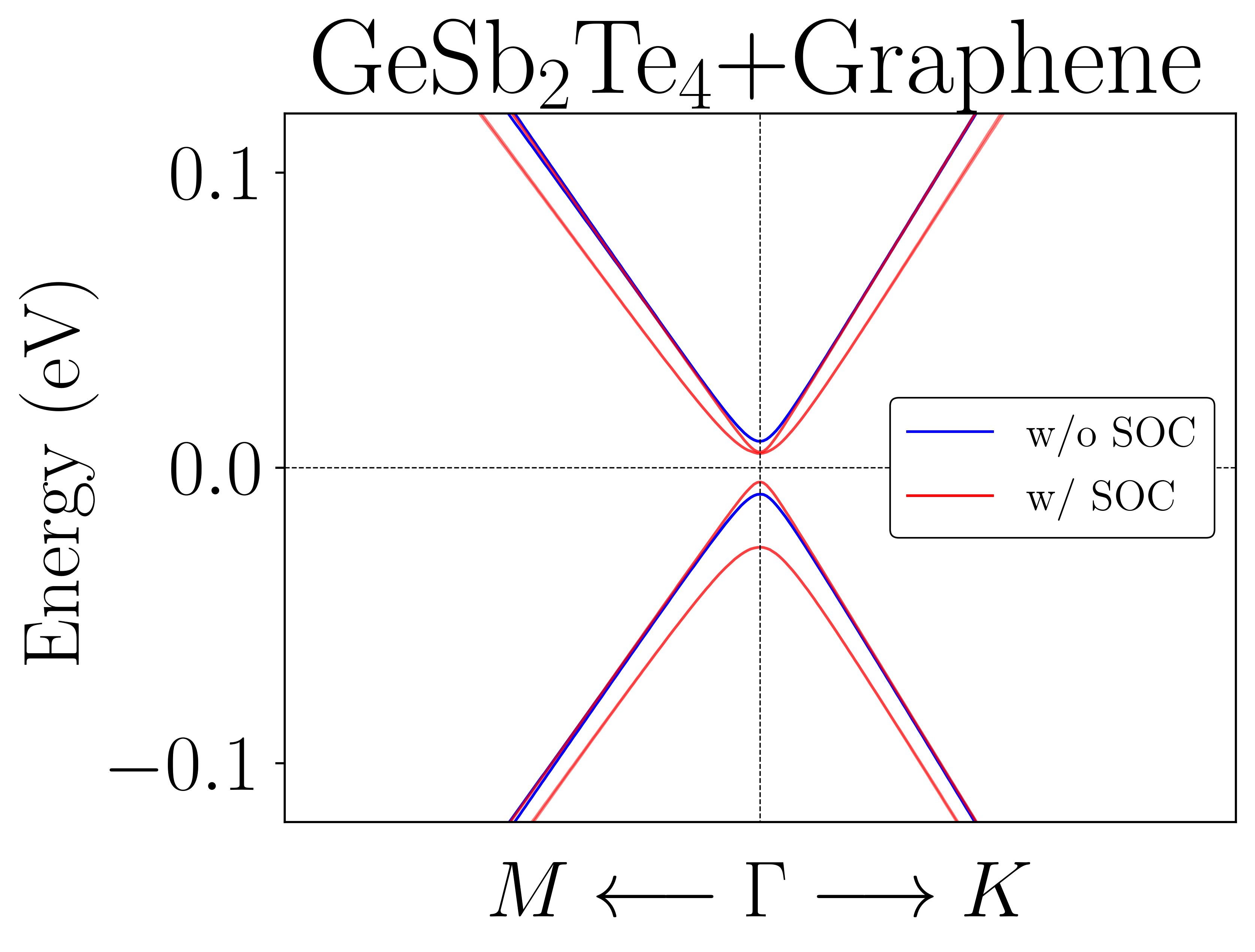}\label{subfig: band structure GeSb2Te4Graphene zoom in}}
\caption{The band structure of a graphene layer on two different monolayer substrates: (a)-(b) for Sb$_2$Te$_3$ and (c)-(d) for GeSb$_2$Te$_4$. The SOC induced by the substrate opens a gap in the graphene, as illustrated in the zoom-in panels (b) and (d).}
\label{fig: band structure substrate}
\end{figure}

From the phase diagrams in the main text (and in \cref{appendix: Additional moir ́e band structures and band topology} for more), we observe topologically non-trivial bands for substrates with a wide range of $m_{xxI}$, $m_{zzz}$, $m_{yyz}$, and $m_{xyz}$. In this section, we study various monolayer substrates with lattice constants that are almost exactly $\sqrt{3}$ times that of graphene, and use density functional theory (DFT) (employing \emph{Quantum Espresso} \cite{Giannozzi_2009, Giannozzi_2017}) to determine the mass terms $m_{xxI}$, $m_{zzz}$, $m_{yyz}$, and $m_{xyz}$.

In practice, we use the Perdew-Burke-Ernzerhof (PBE) exchange-correlation functional, and account for core electrons using Kresse-Joubert projector augmented-wave pseudopotentials. Before performing the band structure calculation, the graphene and the monolayer substrate are relaxed until the residual force on each atom is less than $10^{-4}$ (a.u.). The van der Waals interactions are applied using Grimme's DFT-D2 method, and a Monkhorst–Pack $k$-point grid of $9\times 9 \times 1$ is employed. The mass term $m_{xxI} \tau_{x}\sigma_{x}$ can be determined by studying the band gap at the $\boldGamma$ point $\Delta_{\boldGamma}$, using pseudopotentials that don't include relativistic effects, where $\Delta_{\boldGamma} = 2 m_{xxI}$. The remaining mass terms $m_{zzz}$, $m_{yyz}$, and $m_{xyz}$ can be obtained from the energy spectrum at the $\boldGamma$ point using pseudopotentials that include relativistic effects.

Two candidate monolayer substrates with lattice constants approximately $\sqrt{3} a_0$ where $a_0 = \SI{2.46}{\angstrom}$ are given in \cref{table: candidate substrate sqrt3}, where we listed the mass terms $m_{xxI}$, $m_{zzz}$, $m_{yyz}$, $m_{xyz}$, the substrate lattice constant $a_s$, the percentage of deviation from $\sqrt{3} a_{0}$, and the spacing between the graphene and (the upper most atom of) the substrate. The band structures of graphene on the substrate are shown in \cref{fig: band structure substrate}. The corresponding mass terms give rise to relatively flat $n=-2$ bands that are isolated from other bands, both having spin Chern numbers $\mathcal{C}=+2$.

To fit the DFT result, we kept only the $s_z$ preserving terms in the Hamiltonian, as discussed in \cref{appendix: symmetries}. It is possible that spin $s_z$ non-conserving terms are present, but we leave their analysis for future work.

\section{Additional moir\'e band structures and band topology}
\label{appendix: Additional moir ́e band structures and band topology}

\subsection{Band structure of $\text{Ge}\text{Sb}_2\text{Te}_4$}

\begin{figure}[h]
\centering
\includegraphics[width=170mm]{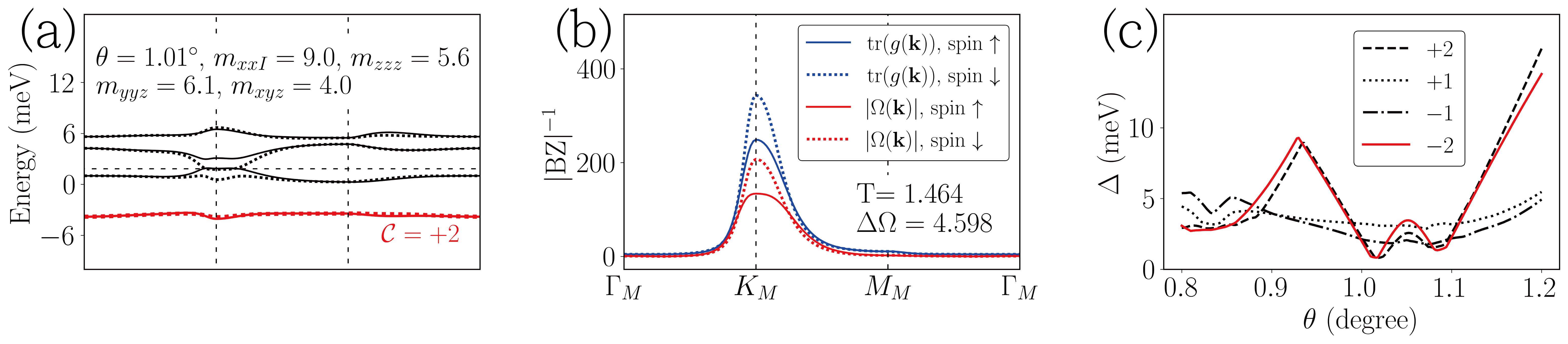}
\caption{(a) The moir\'e band structure for $\text{Ge}\text{Sb}_2\text{Te}_4$, another candidate type-Y substrate. (b) $\tr(g(\boldk))$ and $|\Omega(\boldk)|$ for band $n=-2$. (c) The bandwidths of band $n$ (labeled in the legend) with respect to twist angle $\theta$.}
\label{fig: information of GeSb2Te4}
\end{figure}

In the main text, we presented the moir\'e band structure with the candidate substrate $\text{Sb}_2\text{Te}_3$. Here, we analyze another candidate substrate, $\text{Ge}\text{Sb}_2\text{Te}_4$, and provide its moir\'e band structure and quantum geometry. \cref{fig: information of GeSb2Te4} shows the $|n|\le2$ moir\'e bands at $\theta=1.01^\circ$, where solid and dotted lines stand for spin $\uparrow$ and $\downarrow$ bands, respectively. The bands $|n|\le2$, ordered from high to low energies, carry spin Chern numbers $\mathcal{C}=\{+2, 0, -4, +2\}$.

\subsection{Bands with $|C|=4$}

\begin{figure}[h]
\centering
\subfigure[]{\includegraphics[width=50mm]{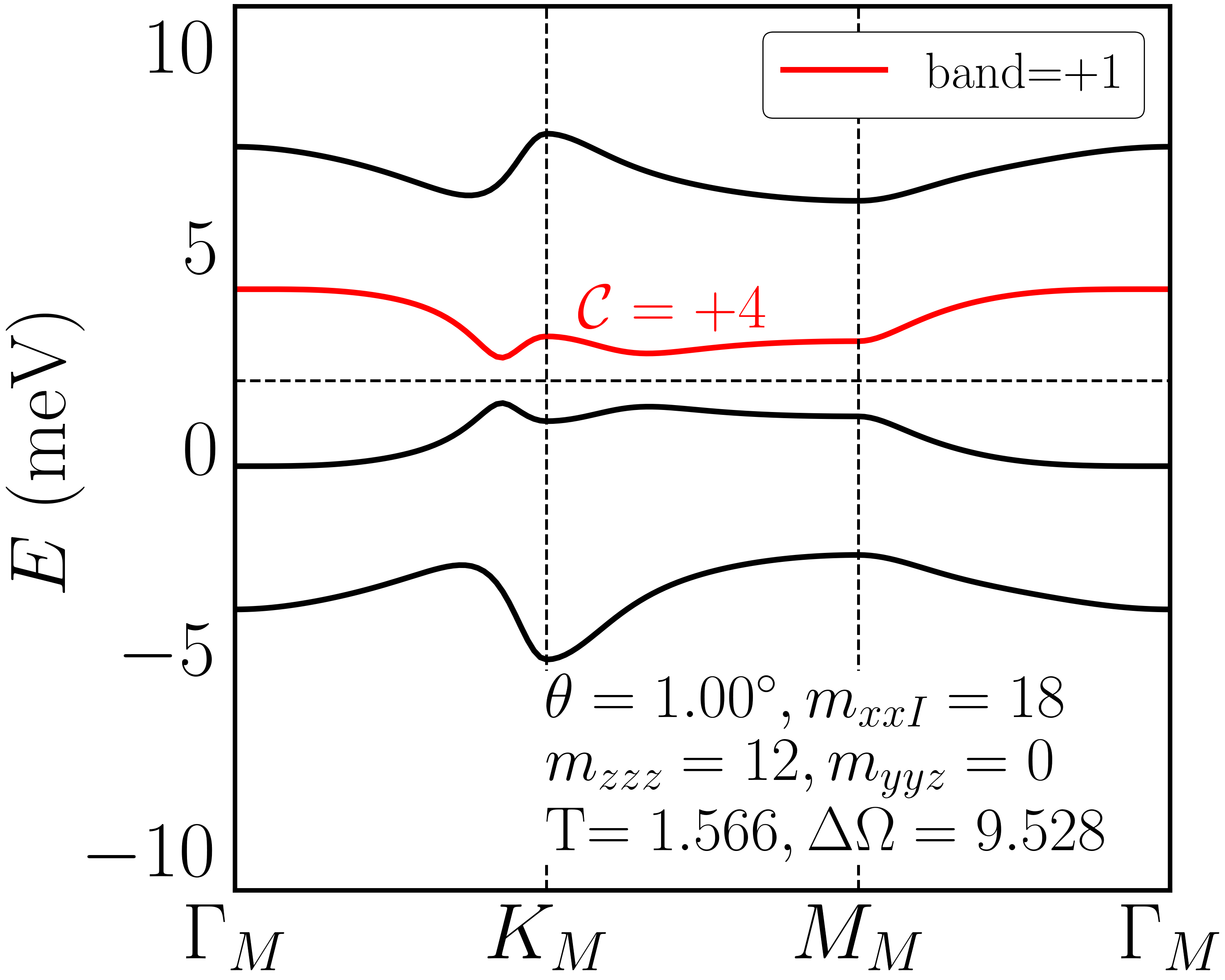}\label{subfig: band structure InfoC4mXXImZZZ}}
\subfigure[]{\includegraphics[width=50mm]{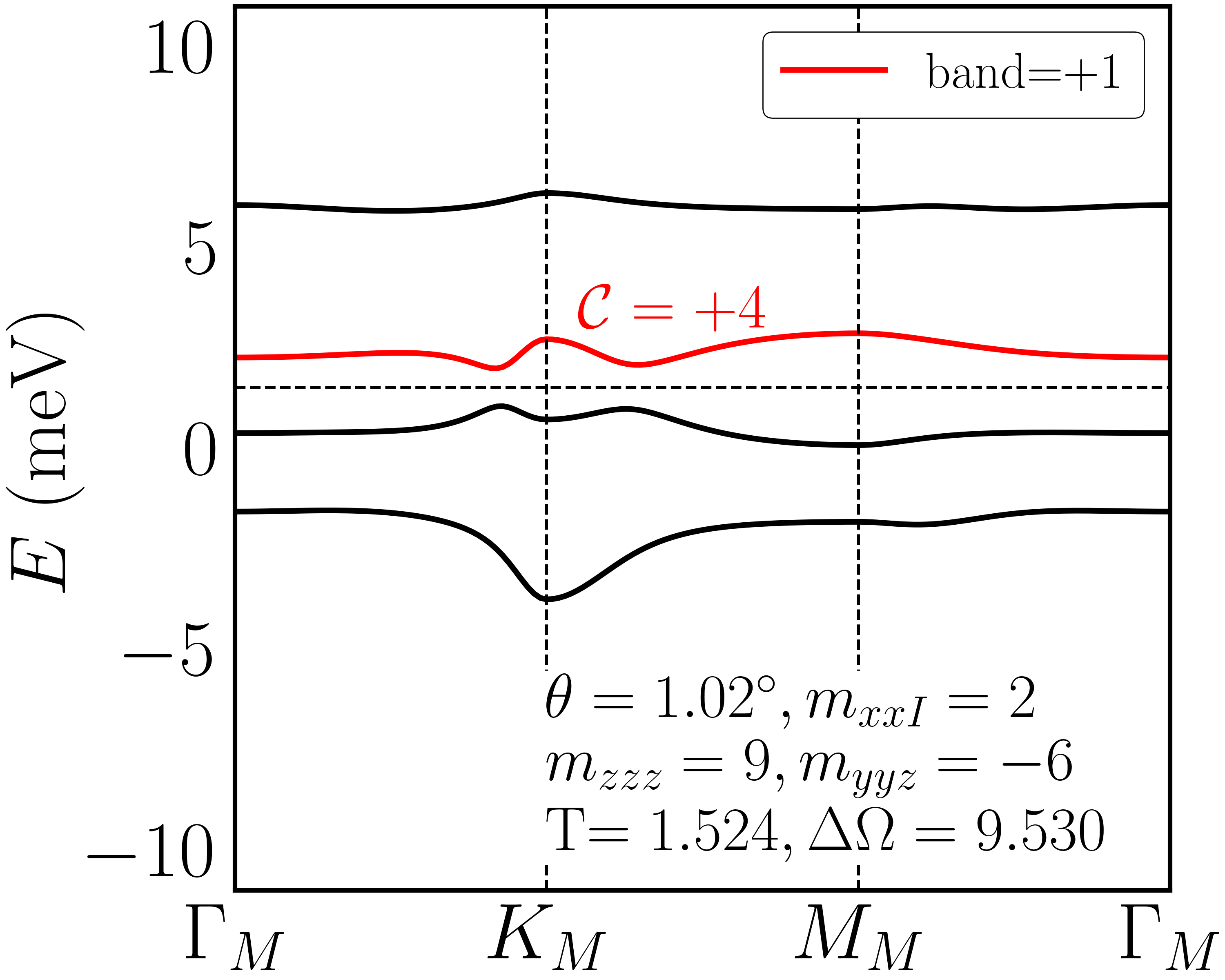}\label{subfig: band structure InfoC4mXXImZZZmYYZ}}
\subfigure[]{\includegraphics[width=50mm]{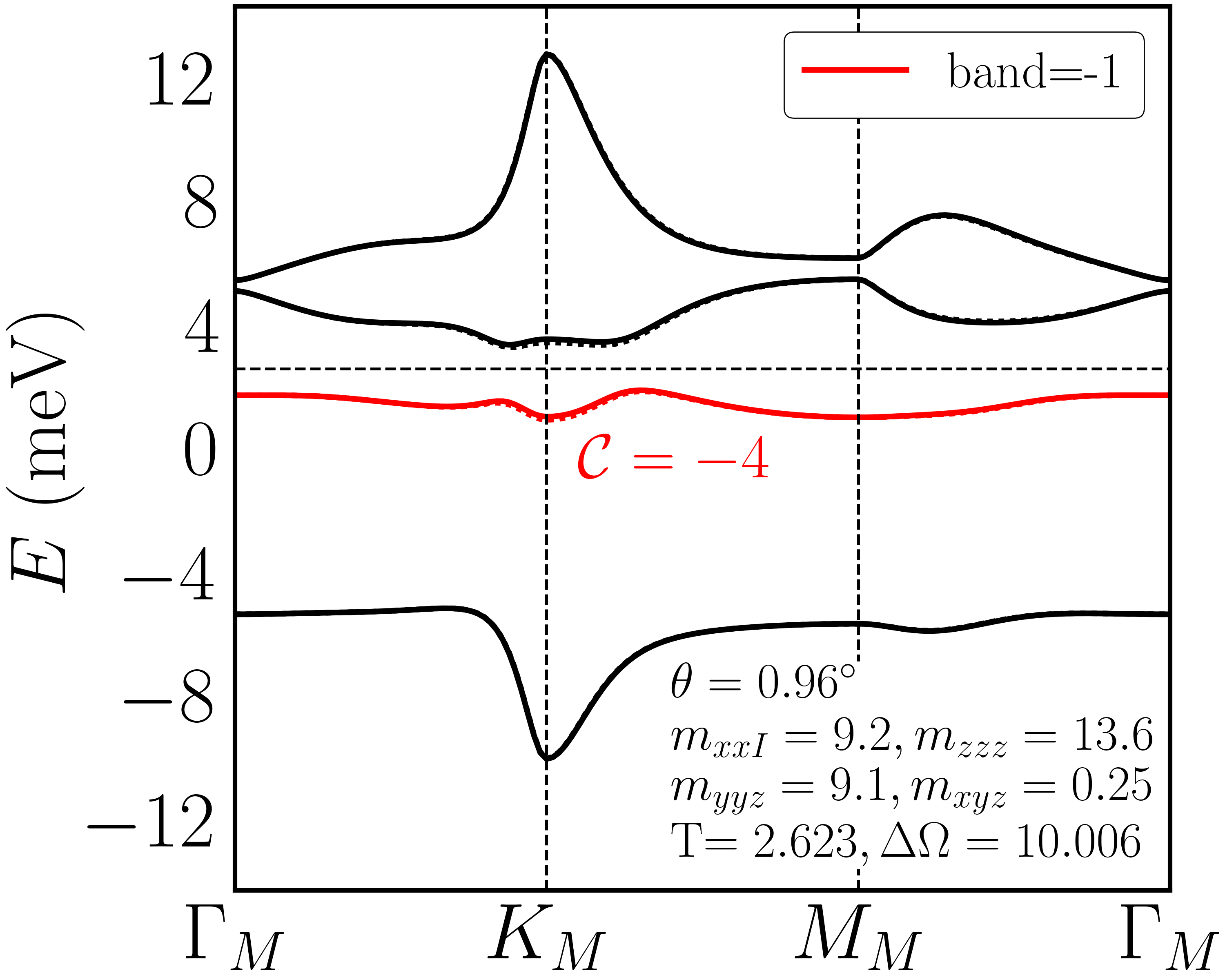}\label{subfig: band structure InfoC4Sb2Te3}}
\caption{The moir\'e band structure for (a) a model with higher spin Chern numbers $\mathcal{C}=\{-2, +4, -4, +2\}$, (b) another model with $\mathcal{C}=\{-2, +4, -4, +2\}$, and (c) the candidate material Sb$_2$Te$_3$, where $\mathcal{C}=\{-1, +3, -4, +2\}$ for bands $n=\{2,1,-1,-2\}$. All masses are given in units of meV.}
\label{fig: information of flat bands C=4}
\end{figure}

In the main text, we showed bands with spin Chern numbers $|\mathcal{C}|=1, 2$. For suitably chosen $\theta$ and mass terms, there also exist isolated flat bands with higher spin Chern numbers. Some examples are shown in \cref{fig: information of flat bands C=4}, including the real material Sb$_2$Te$_3$ (\cref{subfig: band structure InfoC4Sb2Te3}), where the red band indicates bands with spin Chern number $|\mathcal{C}|=4$. The band geometry properties $\text{T}$ and $\Delta\Omega$ are indicated. Notably, the deviations from the ideal band $\text{T}$ for these $|\mathcal{C}|=4$ bands are greater than the main text examples with $|\mathcal{C}|=1, 2$. In addition, the band gaps between these topological flat bands and their neighboring bands are much smaller than the band gaps shown in the main text, making them more challenging to observe experimentally.

\subsection{Band topology}

\begin{figure}[h]
\centering
\includegraphics[width=135mm]{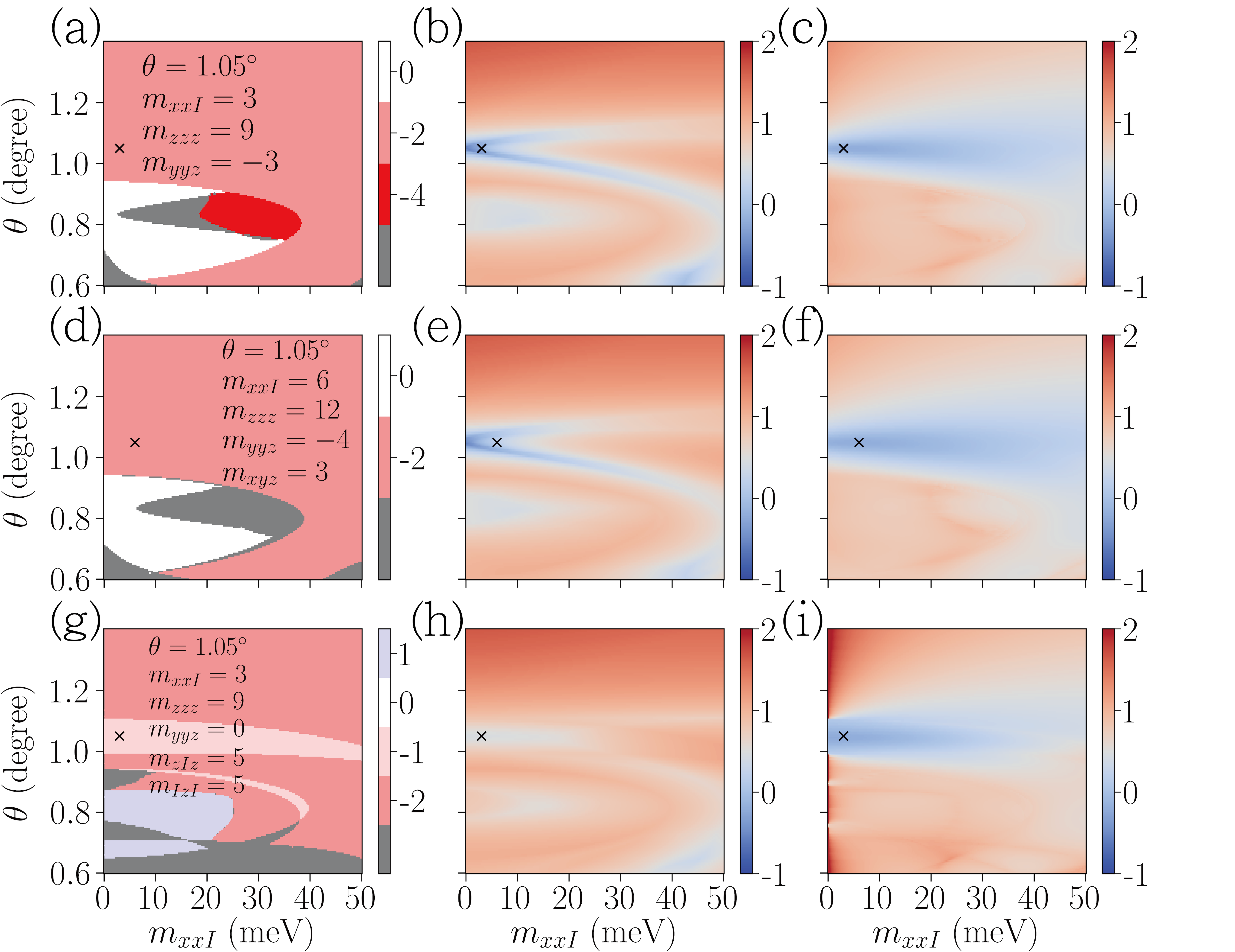}
\caption{The phase diagram of the spin Chern number, band gap $\Delta$, and the $\text{T}$ value for models with (a)-(c) a maximally symmetry substrate, (d)-(f) a type-Y substrate, and (g)-(i) a type-X substrate. Diagrams in each row correspond to the same set of mass terms, as indicated in the left-most panel. The highlighted points in panel (a)-(c) and (g)-(i) correspond to parameters that are described in main text Fig. 2(f) and Fig. 3(d).}
\label{fig: additional phase diagrams SM}
\end{figure}

In the main text, we showed the spin Chern number of models with $m_{zzz}=9$ over a wide range of $\theta$ and $m_{xxI}$. In \cref{fig: additional phase diagrams SM}, we provide additional phase diagrams for various models. The specific details of each diagram are indicated.

\begin{figure}[h]
\centering
\subfigure[]{\includegraphics[width=65mm]{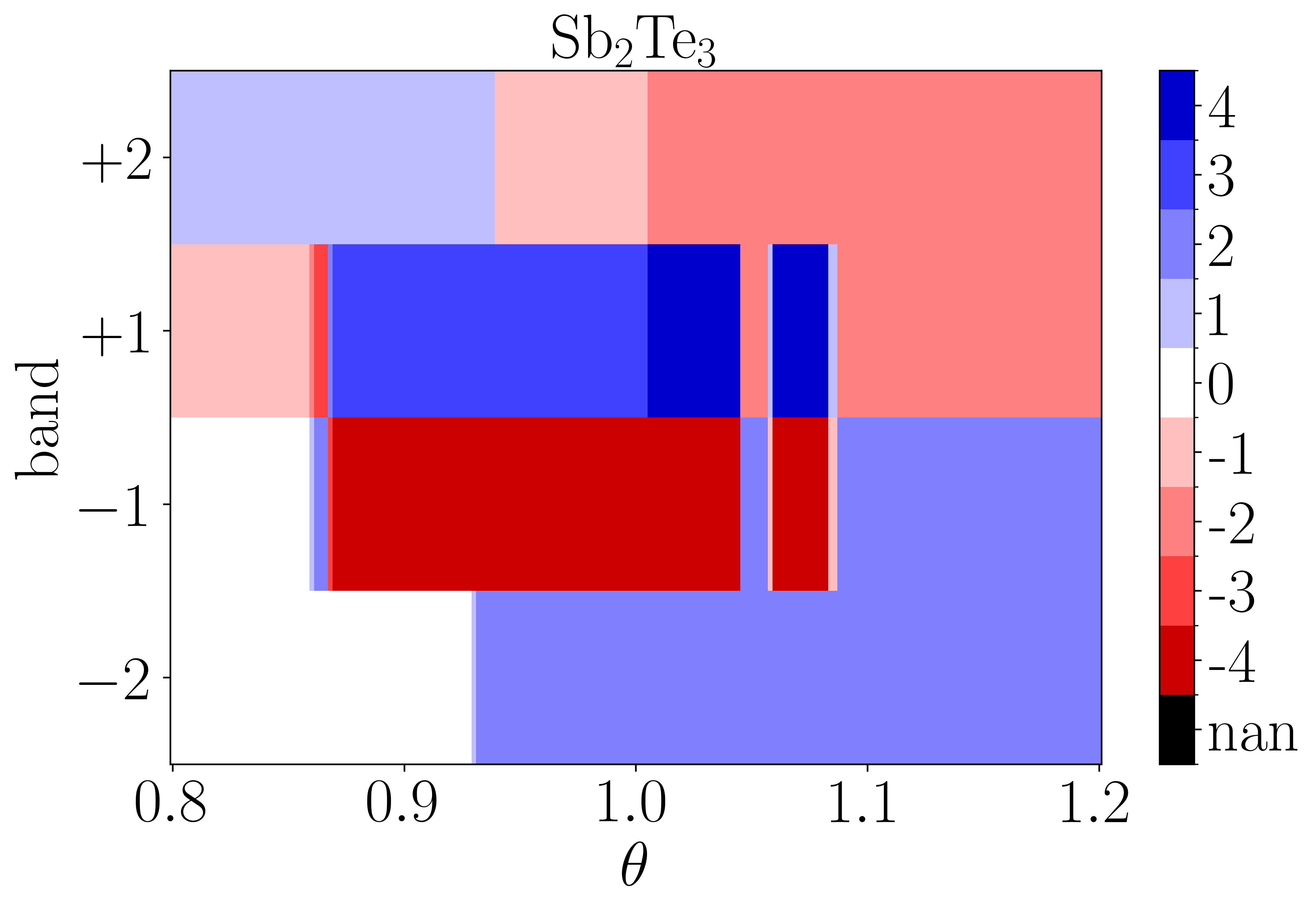}\label{subfig: sb2te3 C angle dependence}}
\subfigure[]{\includegraphics[width=65mm]{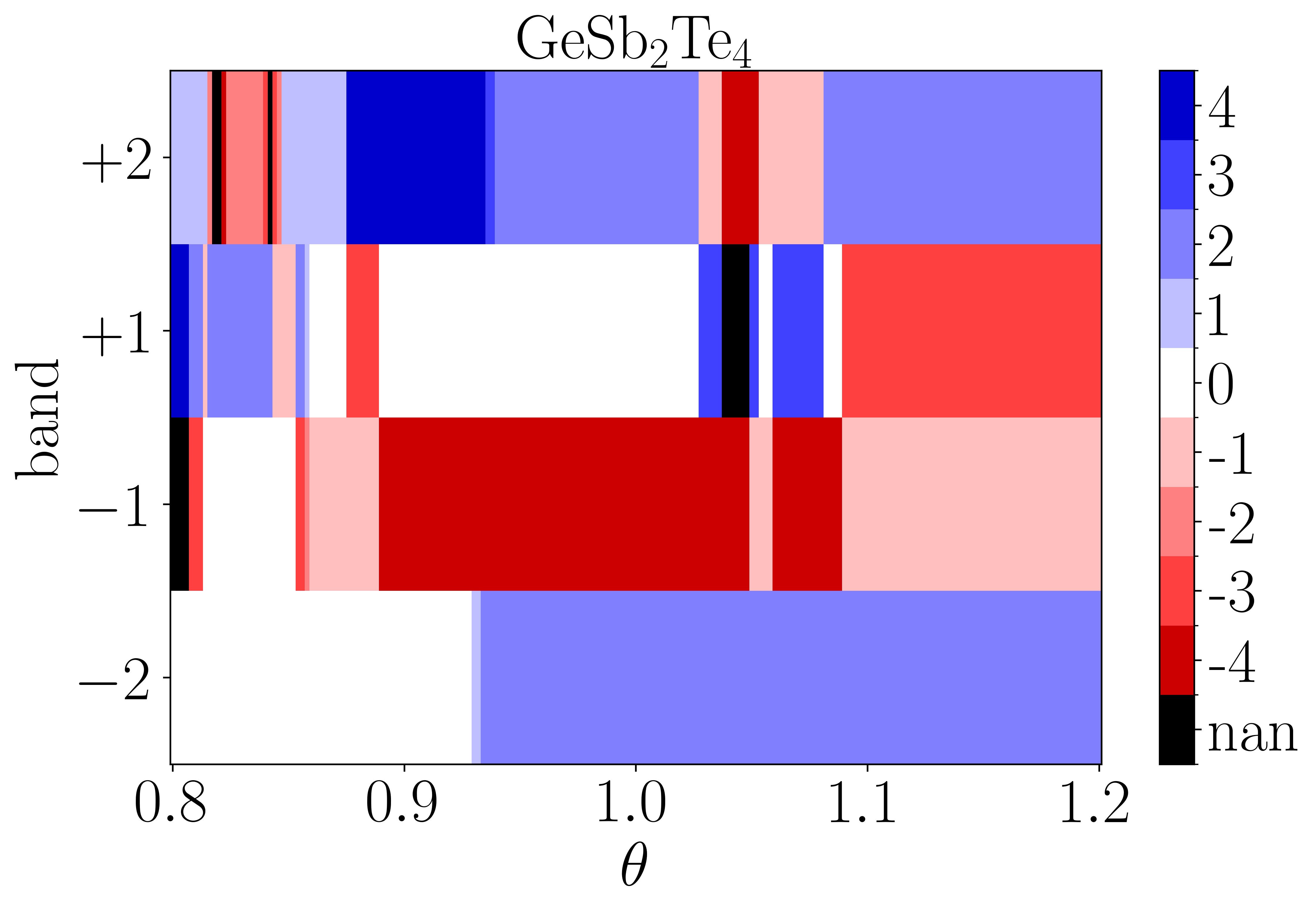}\label{subfig: gesb2te4 C angle dependence}}
\caption{The dependence of the spin Chern number on the twist angle of TBG on (a) Sb$_2$Te$_3$ and (b) GeSb$_2$Te$_4$. Note that for certain values of $\theta$, the total spin Chern number of the four bands is nonzero. For example, in Sb$_2$Te$_3$, this occurs for $\theta \in [0.8, 0.93]$. This happens because bands $n=-2$ and $n=-3$ touch at $\theta \approx 0.93$, making it necessary to include additional bands in order for the total Chern number to vanish. For certain ranges of $\theta$, the spin Chern number can reach values as large as $|\mathcal{C}| > 4$. However, the corresponding band gaps are very small in these regimes and therefore of little experimental interest, so we indicate them in black.}
\label{fig: spin Chern number and angle}
\end{figure}

In \cref{fig: spin Chern number and angle}, we show the spin Chern number of the two candidate substrates across a wide range of twist angles, demonstrating the robustness of the topological phase.

\section{Wilson loop and the quantum geometric tensor}\label{appendix: Wilson loop and quantum geometry tensor}
The Berry curvature is a quantity that captures the topological properties of isolated bands. Recently, attention has also turned to a closely related quantity called the quantum geometric tensor (QGT). In this section, we briefly introduce the definition the QGT. For an isolated band $\ket{\psi(\boldk)}$, the QGT is defined by
\begin{equation}
\label{eqn: quantum geometry tensor appendix}
G_{ij}(\boldk) = 
\tr\left[ 
P(\boldk) \partial_{k_i} P(\boldk) \partial_{k_j} P(\boldk) 
\right]
= g_{ij}(\boldk) + \frac{i}{2} f_{ij}(\boldk),
\end{equation}
where $P(\boldk) = \ket{\psi(\boldk)} \bra{\psi(\boldk)}$, $g_{ij}(\boldk)$ is a real symmetric matrix called the Fubini-Study metric (FSM) \cite{Fubini1904, Study1905, yu2024quantumgeometryquantummaterials}, and $f_{ij}(\boldk)$ is a real antisymmetric matrix, where $\Omega(\boldk) = -f_{xy}(\boldk)$ is the Berry curvature. In the following, we derive useful formulas for $g_{ij}(\boldk)$ and $\Omega(\boldk)$.

We define the overlap function $w(\boldk_1, \boldk_2) = \braket{\psi(\boldk_1)|\psi(\boldk_2)}$. Taking $\boldk_1 = \boldk - \delta\boldk/2$ and $\boldk_2 = \boldk + \delta\boldk/2$ the overlap function can be expanded to second order in $\delta\boldk$ as follows:
\begin{equation}\label{eqn: expansion of inner product}
\begin{split}
& w \left(\boldk-\frac{\delta \boldk}{2}, \boldk + \frac{\delta \boldk}{2} \right)\\
= & 
\braket{\psi(\boldk-\delta\boldk/2) | \psi(\boldk+\delta\boldk/2)}\\
= &
\left(
\bra{\psi(\boldk)}
- \frac{1}{2} \delta k^i \bra{\partial_{k_i}\psi(\boldk)} 
+ \frac{1}{8} \delta k^i \delta k^j \bra{\partial_{k_i}\partial_{k_j}\psi(\boldk)} 
\right)
\left(
\ket{\psi(\boldk)}
+ \frac{1}{2} \delta k^i \ket{\partial_{k_i}\psi(\boldk)} 
+ \frac{1}{8} \delta k^i \delta k^j \ket{\partial_{k_i}\partial_{k_j}\psi(\boldk)} 
\right)
\\
= &
1 
+ \delta k^i \braket{\psi(\boldk) | \partial_{k_i} \psi(\boldk)}
- \frac{1}{2} \delta k^i \delta k^j  \braket{\partial_{k_i}\psi(\boldk) | \partial_{k_j} \psi(\boldk)},
\end{split}
\end{equation}
where repeated indices are summed over. Note that we have used $\braket{\psi(\boldk) | \partial_{k_i} \psi(\boldk)}+\braket{\partial_{k_i} \psi(\boldk) | \psi(\boldk)}=0$ and $\braket{\psi(\boldk) | \partial_{k_i} \partial_{k_j} \psi(\boldk)}+\braket{\partial_{k_i}\partial_{k_j} \psi(\boldk) | \psi(\boldk)}=- \braket{\partial_{k_i} \psi(\boldk) | \partial_{k_j} \psi(\boldk)} - \braket{\partial_{k_j} \psi(\boldk) | \partial_{k_i} \psi(\boldk)}$. On the other hand, the Taylor expansion of an exponential function with arbitrary coefficients $\alpha_i$ and $\beta_{ij}$ to second order in $\delta\boldk$ is
\begin{equation}\label{eqn: expansion of exponential function}
\exp\left( \alpha_i \delta k^i + \beta_{ij} \delta k^i \delta k^j \right)
=
1 + \alpha_i \delta k^i + \left( \frac{1}{2} \alpha_i \alpha_j + \beta_{ij} \right) \delta k^i \delta k^j.
\end{equation}
By comparing \cref{eqn: expansion of inner product,eqn: expansion of exponential function} and setting $\alpha_i 
= \braket{\psi(\boldk) | \partial_{k_i} \psi(\boldk)} 
= -i \mathcal{A}_i(\boldk)$ and $\frac{1}{2} \alpha_i \alpha_j + \beta_{ij} = - \frac{1}{2} \braket{\partial_{k_i}\psi(\boldk) | \partial_{k_j} \psi(\boldk)}$, we obtain
\begin{equation}
\begin{split}
w \left(
\boldk-\frac{\delta \boldk}{2}, \boldk + \frac{\delta \boldk}{2} 
\right) 
= &
\exp\left[ 
-i \mathcal{A}_i(\boldk) \delta k^i - \frac{1}{2} \braket{\partial_{k_i}\psi(\boldk) | \partial_{k_j} \psi(\boldk)}\delta k^i\delta k^j + \frac{1}{2} \mathcal{A}_i(\boldk) \mathcal{A}_j(\boldk) \delta k^i\delta k^j
\right]
\\
= & 
\exp\left[ 
-i \mathcal{A}_i(\boldk) \delta k^i - \frac{1}{2} \hat{g}_{ij}(\boldk) \delta k^i\delta k^j
\right],
\end{split}
\end{equation}
to second order in $\delta\boldk$, where $\hat{g}_{ij}(\boldk) = \text{Re}\left[ \braket{\partial_{k_i}\psi(\boldk) | \partial_{k_j} \psi(\boldk)} \right] - \mathcal{A}_i(\boldk) \mathcal{A}_j(\boldk)$.

Given a closed loop $\Gamma$ formed by line segments between $N$ discrete momenta $\boldk_{1}, \boldk_{2}, \cdots, \boldk_{N}, \boldk_{N+1}=\boldk_{1}$, the gauge-invariant Wilson loop unitary is
\begin{equation}\label{eqn: W in terms of Omega and g}
\begin{split}
W(\boldk_1, \boldk_2, \cdots, \boldk_N) =& \prod_{n=1}^{N} w(\boldk_{n}, \boldk_{n+1})
\\
=& 
\exp\left[ 
-i \sum_n \mathcal{A}_i(\bar{\boldk}_n) \delta k_n^i - \frac{1}{2} \sum_n \hat{g}_{ij}(\bar{\boldk}_n) \delta k_n^i\delta k_n^j
\right]
+\mathcal{O}(|\delta \boldk|^3)
\\
=& 
\exp\left[ 
-i \int_\Gamma \hat{\Omega}(\boldk) d^2\boldk - \frac{1}{2} \sum_n \hat{g}_{ij}(\bar{\boldk}_n) \delta k_n^i\delta k_n^j
\right]
+\mathcal{O}(|\delta \boldk|^3),
\end{split}
\end{equation}
where $\delta \boldk_n = \boldk_{n+1} - \boldk_{n}$, and $\bar{\boldk}_n = \frac{1}{2} ( \boldk_{n+1} + \boldk_{n} )$ and
\begin{equation}
\hat{\Omega}(\boldk) = \partial_{k_x} \mathcal{A}_y(\boldk) - \partial_{k_y} \mathcal{A}_x(\boldk).
\end{equation}

We now prove $g_{ij}(\boldk) = \hat{g}_{ij}(\boldk)$ and $\Omega(\boldk) = \hat{\Omega}(\boldk)$.
\begin{itemize}
    \item $g_{ij}(\boldk)=\hat{g}_{ij}(\boldk)$:
    Using the identities $\tr(A \{ B, C \}) = \tr(B \{ A, C \})$ and $\{ P(\boldk), \partial_{k_i} P(\boldk) \} = \partial_{k_i} P(\boldk)$, and referring to the QGT defined in \cref{eqn: quantum geometry tensor appendix}, we obtain
    \begin{equation}
        \begin{split}
            g_{ij}(\boldk)=&
            \frac{1}{2}
            \tr\left[ 
            P(\boldk) \{ \partial_{k_i} P(\boldk), \partial_{k_j}  P(\boldk) \}
            \right]
            \\
            =& \frac{1}{2}
            \tr\left[ 
            \partial_{k_i} P(\boldk) \{ P(\boldk), \partial_{k_j}  P(\boldk) \}
            \right]
            \\
            =& \frac{1}{2}
            \tr\left[ 
            \partial_{k_i} P(\boldk) \partial_{k_j}  P(\boldk)
            \right]
            \\
            =& \frac{1}{2}\left[ 
            \braket{\partial_{k_i}\psi(\boldk) | \partial_{k_j} \psi(\boldk)} 
            + \braket{\partial_{k_j}\psi(\boldk) | \partial_{k_i} \psi(\boldk)}
            \right] 
            + \braket{\psi(\boldk) | \partial_{k_i} \psi(\boldk)}
            \braket{\psi(\boldk) | \partial_{k_j} \psi(\boldk)}
            \\
            =& \hat{g}_{ij}(\boldk)
            ,
        \end{split}
    \end{equation}
    where, in the last line, we used the fact that $\braket{\psi(\boldk) | \partial_{k_i} \psi(\boldk)}$ is purely imaginary.
    \item $\Omega(\boldk) = \hat{\Omega}(\boldk)$:
    Using the fact that $\braket{\psi | \partial_{k_i}\psi}$ is imaginary, we obtain
    \begin{equation}
    \tr(P \partial_{k_x} P \partial_{k_y} P) = \braket{\partial_{k_x} \psi | \partial_{k_y} \psi} + \braket{\psi | \partial_{k_x} \psi} \braket{\psi | \partial_{k_y} \psi}.
    \end{equation}
    Substituting this into $f_{xy}$, we get
    \begin{equation}
    \begin{split}
    \Omega(\boldk) &= -f_{xy} (\boldk)\\
    &= i \tr(P [\partial_{k_x} P, \partial_{k_y} P ])\\
    &= i (\braket{\partial_{k_x} \psi(\boldk) | \partial_{k_y} \psi(\boldk)} - \braket{\partial_{k_y} \psi(\boldk) | \partial_{k_x} \psi(\boldk)})\\
    &= \hat{\Omega}(\boldk).
    \end{split}
    \end{equation}
\end{itemize}
We have seen that the off-diagonal terms of the QGT capture the first order expansion of the Wilson loop unitary, while the diagonal terms capture the second order expansion of the Wilson loop unitary.

\section{Numerical estimation of the Berry curvature and the quantum geometry}
\label{appendix: Numerical estimation of the Berry curvature and the quantum geometry}

\begin{figure}[h]
\centering
\begin{subfigure}[] {
\centering
    \begin{tikzpicture}
            \def\sqrtThree{1.73205080757}
            \def\side{4}
            \def\Nmesh{8}
            \coordinate (A) at (0, 0);
            \coordinate (B) at (\side, 0);
            \coordinate (C) at (3*\side/2, \sqrtThree*\side/2);
            \coordinate (D) at (1*\side/2, \sqrtThree*\side/2);
        
            \draw[->][line width=0.8mm][red] (-0.2*2/\sqrtThree, 0) -- (1*\side/2-0.2*2/\sqrtThree, \sqrtThree*\side/2);
            \draw[->][line width=0.8mm][red] (0, -0.2) -- (\side, -0.2);
            \node[font=\Large][red] at (\side*0.05, \side*0.5) {$\vec{G}_2$};
            \node[font=\Large][red] at (\side*0.5, -\side*0.15) {$\vec{G}_1$};
        
            \draw[line width=0.5mm] (A) -- (B) -- (C) -- (D) -- cycle;
        
            \foreach \i in {1,2,...,\numexpr\Nmesh-1\relax} {
                \draw[gray] (0+\i*\side/\Nmesh, 0) -- (\side/2+\i*\side/\Nmesh, \sqrtThree*\side/2);
                \draw[gray] (0+\i*\side/\Nmesh/2, \sqrtThree*\side*\i/2/\Nmesh) -- (\side+\i*\side/\Nmesh/2, \sqrtThree*\side*\i/2/\Nmesh);
            }
        
            \foreach \i in {0,1,...,\numexpr\Nmesh\relax} {
                \foreach \j in {0,1,...,\numexpr\Nmesh\relax} {
                    \fill[blue] (\i*\side/\Nmesh+\j*\side/2/\Nmesh, \j*\sqrtThree*\side/\Nmesh/2) circle (2 pt);
                }
            }
    \end{tikzpicture}
}
\label{subfig: BZ with mesh}
\end{subfigure}
\hspace{10pt}
\begin{subfigure}[]{
\centering
    \begin{tikzpicture}
            \def\sqrtThree{1.73205080757}
            \definecolor{darkgreen}{rgb}{0.0, 0.5, 0.0}
            \def\side{3}
            \def\Nmesh{8}
            \coordinate (A) at (0, 0);
            \coordinate (B) at (\side, 0);
            \coordinate (C) at (3*\side/2, \sqrtThree*\side/2);
            \coordinate (D) at (1*\side/2, \sqrtThree*\side/2);
            \coordinate (midAD) at (1*\side/4, \sqrtThree*\side/4);
            \coordinate (midAB) at (1*\side/2, 0);
            \coordinate (midBC) at (5*\side/4, \sqrtThree*\side/4);
            \coordinate (midCD) at (1*\side, \sqrtThree*\side/2);

            \node[font=\large] at (\side*0.5, -\side*0.15) {$\xi=\frac{|\vec{G}|}{N}$};

            \node[font=\large][blue] at (-0.15*\side, -0.12*\side) {$\ket{\psi_A}$};
            \node[font=\large][blue] at (\side+0.15*\side, -0.12*\side) {$\ket{\psi_B}$};
            \node[font=\large][blue] at (3*\side/2+0.15*\side, \sqrtThree*\side/2+0.12*\side) {$\ket{\psi_C}$};
            \node[font=\large][blue] at (1*\side/2-0.15*\side, \sqrtThree*\side/2+0.12*\side) {$\ket{\psi_D}$};

            \draw[line width=0.5mm] (A) -- (B) -- (C) -- (D) -- cycle;
            
            \fill[blue] (A) circle (4 pt);
            \fill[blue] (B) circle (4 pt);
            \fill[blue] (C) circle (4 pt);
            \fill[blue] (D) circle (4 pt);
            \fill[darkgreen] (midAD) circle (4 pt);
            \fill[darkgreen] (midAB) circle (4 pt);
            \fill[darkgreen] (midBC) circle (4 pt);
            \fill[darkgreen] (midCD) circle (4 pt);
            \fill (3*\side/4, \sqrtThree*\side/4) circle (3 pt);
            \node[font=\large] at (3*\side/4-0.1*\side, \sqrtThree*\side/4+0.1*\side) {$P$};
    \end{tikzpicture}
}
\label{subfig: a single mesh of BZ}
\end{subfigure}
\caption{(a) The rhombus spanned by $\vec{G}_1$ and $\vec{G}_2$ is divided into $N \times N$ cell, each having side length $\xi=|\vec{G}|/N$. (b) Focusing on a particular cell, the center is labeled by $P$, while the corners are labeled $A$, $B$, $C$, and $D$, where the wavefunctions are $\ket{\psi_A}$, ..., $\ket{\psi_D}$.}
\label{fig: BZ mesh}
\end{figure}

In this section, we introduce a gauge invariant numerical method for calculating $\Omega(\boldk)$ and $g(\boldk) = \tr[ g_{ij}(\boldk) ]$ using the same set of sampled points in the BZ. This method applies as long as the reciprocal lattice is spanned by primitive vectors $G_1$ and $G_2$ with $|G_1| = |G_2|$ and an angle of $\pi/3$ between $G_1$ and $G_2$, as shown in \cref{fig: BZ mesh}(a). Importantly, these conditions can be met for the model studied in this paper.

We start by dividing the BZ into an $N \times N$ grid, as illustrated in \cref{fig: BZ mesh}(a). The nodes, indicated by blue points, are the sampled momenta at which we diagonalize the Hamiltonian $H(\boldk)$ to obtain the eigenvectors and eigenvalues. A small grid section is shown in \cref{fig: BZ mesh}(b), which is a rhombus with sides of length $\xi = |G_1|/N = |G_2|/N$. The center of the rhombus is labeled by $P$, while the corners are labeled $A$, $B$, $C$, and $D$, and the corresponding wavefunctions for an occupied band are denoted as $\ket{\psi_A}$, ..., $\ket{\psi_D}$. The algorithms for calculating $\Omega(\boldk)$ and $g(\boldk)$ at $P$, denoted by $\Omega(P)$ and $g(P)$, are:
\begin{itemize}
    \item $\Omega(P)$: We can approximate $\Omega(P)$ by the average of $\Omega(\boldk)$ over the grid with area $\sqrt{3}\xi^2 / 2$, with the deviation up to order $\mathcal{O}(\xi^2)$.
    According to \cref{eqn: W in terms of Omega and g}, we have
    \begin{equation}
        \frac{\sqrt{3}}{2} \xi^2 \Omega(P)
        \approx
        \int \Omega(\boldk) d^2 \boldk
        = -\text{Im} \left[ \log W(\boldk_A, \boldk_B, \boldk_C, \boldk_D) \right]
    .\end{equation}
    Note that this equation can only determine $\frac{\sqrt{3}}{2} \xi^2 \Omega(P)$ up to a modulo of $1$. However, if $\xi$ is sufficiently small, the LHS of the equation becomes much less than $1$, allowing us to disregard the ambiguity of integer shifts.
    \item $g(P)$:
    From the vectors $\overrightarrow{AC}=\left( \frac{3}{2}\xi, \frac{\sqrt{3}}{2}\xi \right)$ and $\overrightarrow{BD}=\left( -\frac{1}{2}\xi, \frac{\sqrt{3}}{2}\xi \right)$, and in accordance with \cref{eqn: W in terms of Omega and g}, we expand the logarithm of the Wilson loop unitary around $P$ to $\mathcal{O}(\xi^2)$ as:
    \begin{equation}
        \begin{aligned}
            - \log W(\boldk_A, \boldk_C) =& 
            g_{xx}(P) |\overrightarrow{AC}_x|^2
            + 2 g_{xy}(P)  (\overrightarrow{AC}_x \cdot \overrightarrow{AC}_y)
            + g_{yy}(P) |\overrightarrow{AC}_y|^2
            \\
            =& 
            \frac{9}{4}\xi^2 g_{xx}(P) 
            + \frac{6\sqrt{3}}{4}\xi^2 g_{xy}(P)
            + \frac{3}{4}\xi^2 g_{yy}(P)
            \\
            - \log W(\boldk_B, \boldk_D) =& 
            \frac{1}{4}\xi^2 g_{xx}(P) 
            - \frac{2\sqrt{3}}{4}\xi^2 g_{xy}(P)
            + \frac{3}{4}\xi^2 g_{yy}(P)
        \end{aligned}
    .\end{equation}
    Combining the equations gives
    \begin{equation}
        - \log W(\boldk_A, \boldk_C) - 3 \log W(\boldk_B, \boldk_D)
        = 3\xi^2 \left( g_{xx}(P) + g_{yy}(P) \right)
        = 3\xi^2 g(P).
    \end{equation}
\end{itemize}

The algorithm provided calculates $\Omega(P)$ and $g(P)$ up to order $\mathcal{O}(\xi^2)$ using only $(N+1)\times(N+1)$ samples. To see its efficiency, we compare it to a more straightforward method: sampling the green points in \cref{subfig: a single mesh of BZ}. This alternative approach requires $2N(N+1)$ samples, which is approximately double the number by our method.

\end{document}